\documentclass[
 preprint,
superscriptaddress,
 amsmath,amssymb,
 aps,
]{revtex4-1}

\usepackage{graphicx}
\usepackage{dcolumn}
\usepackage{bm}
\usepackage{hyperref}

\begin{document}

\preprint{LMU-ASC 18/16}

\title{Skyrmion Black Hole Hair:  Conservation of Baryon Number by Black Holes  and Observable Manifestations
}

\author{Gia Dvali}
\affiliation{%
 Arnold Sommerfeld Center, Ludwig-Maximilians-Universit\"at, 80333 M\"unchen, Germany\\
}
\affiliation{%
 Max-Planck-Institut f\"ur Physik, Werner-Heisenberg-Institut, 80805 M\"unchen, Germany\\
}
\affiliation{%
 Center for Cosmology and Particle Physics, Department of Physics, New York University, 4 Washington Place, New York, NY 10003, USA\\
}

\author{Alexander Gu\ss mann}
 \email{alexander.gussmann@physik.uni-muenchen.de}
\affiliation{%
 Arnold Sommerfeld Center, Ludwig-Maximilians-Universit\"at, 80333 M\"unchen, Germany\\
}

\date{\today}

\begin{abstract}
We show that the existence of black holes with classical skyrmion hair invalidates standard proofs that global charges, such as the baryon number, cannot be conserved by a black hole.  By carefully analyzing the standard arguments based on a Gedankenexperiment in which a black hole is seemingly-unable to return  the baryon number that it swallowed,   
we identify  inconsistencies in this reasoning,  which does not take into the account neither the existence of skyrmion black holes nor the 
 baryon/skyrmion correspondence.  We then perform a refined  Gedankenexperiment by incorporating the new knowledge and show that no contradiction with conservation of baryon number takes place at any stage of black hole evolution. 
 Our analysis also indicates no conflict between semi-classical black holes and the existence of baryonic gauge interaction  
 arbitrarily-weaker than gravity.  
Next, we study classical cross sections of a minimally-coupled massless probe scalar field scattered by a skyrmion black hole. We investigate how the skyrmion hair manifests itself  by comparing this cross section with the analogous cross section 
caused by a Schwarzschild black hole which has the same ADM mass as the skyrmion black hole. 
  Here we find an order-one difference in the positions of the characteristic peaks in the cross sections. The peaks are shifted to smaller scattering angles when the skyrmion hair is present. This comes from the fact that the skyrmion  hair changes the near horizon geometry of the black hole when compared to a Schwarzschild black hole with same ADM mass. We keep the study of this second aspect general so that the qualitative results which we obtain can also be applied to black holes with classical hair of different kind.
\end{abstract}

\maketitle

\section{\label{sec:level1}Introduction\protect\\}
In 1971 Ruffini and Wheeler noted that a black hole formed by gravitational collapse is completely described by its mass, electric charge and angular momentum and by no other additional parameters, such as, baryon number or strangeness \cite{wheeler}. They emphasized that electric charge compared to other physical parameters, like baryon number, is distinguished in the sense that it is associated to a Gauss law.  Therefore, the association of a Gauss law to a physical parameter can - in the spirit of Ruffini and Wheeler - be conjectured to be a necessary and sufficient condition for the parameter to survive a gravitational collapse in the sense that it appears for an external observer as one of the parameters characterizing the resulting black hole. This is one formulation of what became well known as the ``no-hair" conjecture. In the following year the first so called ``no-hair" theorems have been proven \cite{beke1, *beke2, *beke3, *teitel1, *teitel2, *hartle}.  They support the no-hair conjecture for some special classical cases with certain types of matter. In other cases classical hairs have been discovered \cite{bbm, *beke4, *gibbons} which albeit do not contradict to the ``no-hair" conjecture in the sense that the hairs are ``secondary", meaning that they can be completely re-parametrized in terms of black hole ``primary"  parameters, which are associated to a Gauss law. However, in 1989 Volkov and Gal'tsov discovered solutions of the Einstein equations with non-abelian matter,  which were interpreted as black holes with classical primary hair (not associated to a Gauss law) \cite{volkov1} and until today many similar solutions of the Einstein equations with different kinds of matter have been found \cite{volkov2, *herdeiro, *vol, *biz, *be}. Many of these black hole solutions turned out to be unstable against spherically-symmetric linear perturbations. However, there are also some black hole solutions of this kind which are known to be stable against such linear perturbations. To our knowledge, all such (spherically-symmetric and asymptotically flat) black hole solutions, which are known to be stable against spherically-symmetric linear perturbations, have in common that the role of the matter source in the Einstein's equations is taken by a topological soliton and the black hole event horizon is located inside the soliton\footnote{These kinds of configurations therefore sometimes also go under the name of \textit{horizons inside classical lumps} \cite{kastor}.}.  Some of the most interesting black holes of this kind are the black holes where the soliton is a skyrmion \cite{luckock, droz1, bizon, shiiki, droz2}. This is because the meson theory, in which skyrmions arise as solitonic configurations, can be considered as a simplified model for a low-energy effective theory of QCD (``simplified" in the sense that the meson degrees of freedom in this low-energy effective theory are restricted to pions) and the topological Chern-Simons current of this low-energy effective theory can be mapped to the baryon number current of QCD \cite{bala, wit1, *wit2}. Therefore, these ``skyrmion black holes" discussed in \cite{luckock, droz1, bizon, shiiki, droz2} are nothing but baryonic black holes with an event horizon which is located inside the baryon, when considered from the point of view of a low-energy observer. 
They serve as the existence proof  of black hole solutions with classical baryon/skyrmion hair. 

Under {\it black holes with classical baryon/skyrmion hair } we mean black hole configurations with non-zero baryon number/skyrme topological charge with a classically-measurable deviation in the field outside of the event horizon (caused by the charge) when compared to a Schwarzschild black hole of the same ADM mass. We cannot exclude the possibility  that the class of solutions of black holes with classical baryon/skyrmion hair is in fact larger than the one covered by  the solutions in \cite{luckock, droz1, bizon, shiiki, droz2}. 
 For the purpose of this paper, the existence of known solutions is enough.   With a larger  class of solutions,  our arguments will only become stronger. 

 In this paper  we shall study two important implications of such a classical skyrmion hair of a black hole. 

First, we shall argue that black holes with classical skyrmion hair provide an important loophole in the proof of a standard ``folk theorem" suggesting that the conserved global charges, 
such as baryon number,  are incompatible with semi-classical black hole  physics. 
 This proof is based on the assumption that a black hole, independently of its initial baryonic charge, evaporates  all the way down to the Planck size without having a chance to return its  baryonic number.  
  The existence of black holes with  skyrmion hair allows us to seriously question the validity of  this assumption.  
    By carefully going through all the logical steps of the same Gedankenexperiment as is used in the  ``proof" of the standard folk theorem,  we show that there is no evidence for a necessary violation of the baryon number
 symmetry.  
  
   Namely, there exists no semi-classical principle that forbids an evaporating black hole - that initially 
 swallowed a baryonic charge - to  undergo a transition at a {\it macroscopic}  size,  during which its 
 baryonic charge re-appears in form of a classical skyrmion hair.   It is crucial, that this critical size, $L$, is not a fixed microscopic length-scale of
 quantum gravity, but rather is a scale of the low-energy theory of skyrmion/baryon and  
is determined by the black hole's baryon number.

  We thus conclude that the seeming inevitable inconsistency between the conservation of the baryon charge and 
  semi-classical black holes is abolished by the existence of  classical skyrmion hair.   This  result shifts the burden of proof of violation of global symmetries 
to a more fundamental theory of quantum gravity.  
   
    Next,  we design two versions of a slightly modified Gedankenexperiment in which, under reasonable assumptions respecting the  basic rules of quantum field theory,  the emergence of the skyrmion classical hair at scale $L$ is the only logical outcome.
    
     In the first version, we represent  the baryon number/skyrme topological charge 
  as a surface integral over a two-sphere at infinity.  
  Under certain assumptions, this surface integral can be 
  measured by an asymptotic observer. This measurement then enables 
  this observer to monitor the baryon/skyrmion charge  without ever entering the region in which the skyrmion and a black hole that swallows it are localized. 
  The resurfacing of the classical skyrmion hair - after the black hole event horizon shrinks to the size $L$ - is then a direct consequence of the conservation of the topological charge.

      In the second version of the modified experiment, in order to monitor the re-emergence of  the black hole baryon number in form of the skyrmion charge, we  introduce a spectator $U(1)$ gauge field $B_{\mu}$ coupled to the baryon number current {\it infinitesimally-weakly}  and then take the limit of zero gauge coupling. 
 The spectator gauge field itself plays no role in black hole dynamics and the existence of black hole skyrmion hair is {\it not}  a
consequence of gauging the baryon number. 
The classical skyrmion hair exists whenever the black hole event horizon is smaller than the scale $L$ of the skyrmion, 
and this fact is totally insensitive to the gauging the baryon number symmetry.  
The sole reason for us to infinitely-weakly gauge the baryon number symmetry,  is to make the information about the baryonic 
hair  classically-accessible for an outside observer at stages when the black hole horizon is larger than the scale $L$ of the skyrmion. 

This information plus the correspondence between the baryon Noether current and the skyrmion topological current \cite{bala, wit1, *wit2} 
enforces the re-emergence of the 
 initial baryonic charge in form of the skyrmion hair,  once the black hole shrinks down to a critical size $L$.    
 
  Another consequence of our analysis is that it also reveals a potential loophole in the argument leading to a so-called {\it weak-gravity conjecture} \cite{weak1, *weak2}, according to which the $U(1)$-gauge coupling cannot be arbitrarily weak, without running into a conflict with black hole physics.  The above-discussed folk theorem on incompatibility  of conserved global charges with the existence of semi-classical black holes can be viewed as the limiting case of 
  the weak-gravity conjecture.   The arguments leading to both of these conjectures are based on the key assumption of 
  non-existence of any sort of a classical hair that stores the global charge that went into a black hole,  as well as, the  
  assumption of thermal evaporation of a black hole before it reaches the Planck length.  
   
     The mapping between the black hole baryonic charge and its classical  skyrmion hair seriously challenges the validity of these assumptions, since it shows that the baryonic charge can  re-surface in form of the skyrmion hair  way before the black hole reaches the Planck length.  As a result, as we shall argue, there is no evidence of any inconsistency between the semi-classical black holes  physics and the existence of either global or  weakly-gauged symmetries.  

 In order to avoid a potential confusion we would like to make it very clear, that we are {\it not} proving that 
baryonic and other global charges are necessarily respected by quantum gravity theory that operates at the Planck length.  There  exist well known examples in which operators violating global symmetries are generated either by short distance physics (e.g., grand unification) or by quantum effects 
(e.g., violation of $B+L$-symmetry in the Standard Model by anomalies).  Obviously,  if operators that violate a given global symmetry, e.g., 
the baryon number,  exist at the level of fundamental quantum theory, a semi-classical black hole will not be able to undo their effect.  In such cases, the baryon number swallowed by black holes can be lost due to its non-conservation at the fundamental level.   
There is nothing inconsistent with such a situation.  What we are showing, however, is that the standard arguments used as the evidence that such violation is imposed on a theory by semi-classical black hole physics 
 contains serious loopholes and cannot be taken as a good argument.  We do not exclude that some more refined argument may exist, but such an argument  will need to eliminate the possibility of any type of  black hole hair that could carry the information about its global charge. 

 Secondly, conservation of baryon number by black holes can have important 
 observational consequences.  For example, the way a black hole influences 
 the surrounding medium can be different from what would be expected in the
 case of a Schwarzschild (or Kerr) black hole, due to the new effect of interaction of external particles with a skyrmion/baryon hair of a black hole.   
  
  As an example of a possible mechanism for such an observational manifestation, in  this paper we study scattering cross-sections of a massless minimally-coupled probe scalar field, which is scattered by a skyrmion black hole. 
   We study the question how the structure of the skyrmion ``hair" manifests itself in such classical scattering 
cross-sections.  For this purpose we investigate the positions of the characteristic ``glory" peaks in the scattering cross-sections. We find out that the positions of the characteristic peaks are shifted to smaller scattering angles when compared to the positions of the peaks in the analogous scattering cross-sections caused by a Schwarzschild black hole with same ADM mass as the skyrmion black hole. We identify that the reason for this shift is that the near-horizon geometry of the black hole is changed by the presence of the skyrmion hair when compared to the near-horizon geometry of a Schwarzschild black hole with the same ADM mass.

Since we do not take into account possible non-gravitational interactions between the skyrmion and the probe scalar field (which may be different in the case of different kinds of classical hair), the qualitative result we obtain - the result that the glory peaks are shifted to smaller scattering angles when the skyrmion hair is present - can also be applied to black holes with different kinds of classical hair whenever this hair influences the near-horizon geometry of the black hole through its interaction with the space-time metric in a similar way as the skyrmion does it in the case of a skyrmion black hole. In this sense the skyrmion black holes should be considered as a working example for such more general studies with different kinds of classical black hole hair.

 Finally, we would like to comment on the relation of our results with the idea that information about
 global charges swallowed by a black hole is encoded in its quantum hair \cite{dvaligomez, *kuhnel}.   
 In this reference, based on a particular microscopic framework \cite{Nportrait, *Nportrait1, Nportrait2, *Nportrait3}
in which a black hole is described as a multi-particle bound-state, it was suggested that the swallowed baryonic charge is not lost and becomes a part of the multi-particle quantum state.
An external observer can in principle read-out this quantum information either by scattering some  
long-wavelength radiation at a black hole or by slowly collecting the information encoded in deviations of  Hawking radiation from the thermal spectrum. 
In general, such an information-recovery process requires some knowledge about the microscopic quantum theory  
of the black hole interior.  This is beyond the focus of the present paper. 

 Instead,  we show that the evidence that a black hole preserves information about its baryonic charge exists 
 already in the semi-classical theory, without the need of knowledge of the quantum sub-structure of a black hole. In other words: \\
 
 {\it The quantum baryonic hair of a black hole has a well-defined classical limit in form of a skyrmion hair.} \\ 
 
  This hair has observational consequences, for example, in scattering of waves, as we shall discuss.  
    Needless to say,  the existence of a
potentially-observable  black hole baryonic hair opens-up an interesting set of new questions for astrophysical searches.

The work is organized as follows: In section II we briefly review the skyrmion black hole solutions with zero pion mass discussed in \cite{luckock, droz1, bizon, shiiki, droz2} and extend these solutions to cases with non-zero pion mass. We give a set of parameters which completely characterizes such a skyrmion black hole. In section III, 
using the knowledge about the existence of black holes with skyrmion hair,  we identify a loophole in the folk theorems which  
state that conserved global charges, such as baryon number,  are inconsistent with black hole physics.  

In section IV we choose two skyrmion black holes corresponding to two explicit points in the parameter space discussed in section II and study classical cross sections of a minimally-coupled massless probe scalar field scattered by these black holes. In section V we compare the scattering cross sections obtained in section IV, with the known scattering cross sections of the same scalar field, considered in section IV, now scattered by Schwarzschild black holes which have the same ADM masses as the skyrmion black holes which were considered in section IV. In Section VI we conclude with a summary and an outlook.

We use units in which the speed of light $c$ and the Boltzmann constant $k_B$ are set equal to one ($c = 1$ and $k_B = 1$), but both the Planck constant $\hbar$ and the Newton constant $G_N$ are kept explicit. For the metric we use the signature $(+,-,-,-)$. We treat skyrmions classically.  So,  strictly speaking, we work with a large number of colors $N_C$ \footnote{This is because - although the mapping between baryons in QCD and skyrmions in the low-energy effective meson theory of QCD itself holds for any number of colors $N_C$ \cite{witten2} - the tree level results in the meson theory give appropriate approximations and the solitons can be described in an adequate way using classical physics only if the coupling constants in the meson theory which scale as $N_C^{-1}$ \cite{thooft1, *thooft2, witten1} are small enough \cite{witten2}.}. We restrict our analysis to the case of two flavors.

\section{\label{sec:level1}Skyrmion black holes\protect\\}

In this section we briefly recapitulate how spherically-symmetric solitonic configurations (``skyrmions") arise formally in the skyrme model and how skyrmion black holes arise if we couple the skyrme Lagrangian to Einstein gravity. For this purpose we provide the setup by giving the skyrme Lagrangian $\mathcal{L}_S$ for the case of two flavors in subsection A. In subsection B we then point out how spherically-symmetric solitonic configurations arise from this setup. Finally, in subsection C, we couple the skyrme Lagrangian to Einstein gravity and review how and in what sense these skyrmions can provide a classical ``hair" of a black hole.

\subsection{The skyrme Lagrangian}
The skyrme Lagrangian $\mathcal{L}_S$ is given by \cite{skyrme1, *skyrme2, adkins1, adkins2}
\begin{equation}
\mathcal{L}_S=\mathcal{L}_2 + \mathcal{L}_4 + \mathcal{L}_m \,,
\label{lagrangian}
\end{equation}
with
\begin{equation}
\mathcal{L}_2= - \frac{F_\pi^2}{4}\mathrm{Tr}\left(U^+\partial_\mu U U^+ \partial^\mu U\right) \,,
\end{equation}
\begin{equation}
\mathcal{L}_4 = \frac{1}{32 e^2} \mathrm{Tr}\left([\partial_\mu U U^+, \partial_\nu U U^+]^2\right) \,,
\end{equation}
\begin{equation}
\mathcal{L}_m = \frac{1}{2} \frac{m_\pi^2}{\hbar^2} F_\pi^2 \left(\mathrm{Tr} U - 2\right) \,,
\end{equation}
where $m_\pi$ is the pion mass and $U$ is a $SU(2)$ matrix which is defined by
\begin{equation}
U = e^{\frac{i}{F_\pi}\pi_a \sigma_a} \,,
\label{matrixU}
\end{equation}
with $\pi_a$ the pion fields and $\sigma_a$ the Pauli matrices. $F_\pi$ is the pion decay constant with dimensionality
\begin{equation}
[F_\pi] = \sqrt{\mathrm{\frac{[mass]}{[length]}}}
\end{equation}
and $e$ is a coupling constant of dimensionality
\begin{equation}
[e]= \frac{1}{\sqrt{\mathrm{[mass][length]}}} \, .
\end{equation}
Using these two parameters $F_\pi$ and $e$ one can define a length-scale $L$ and a mass-scale $M_S$ as
\begin{equation}
L = \frac{1}{e F_\pi} \,, 
\end{equation}
\begin{equation}
M_S = \frac{F_\pi}{e} \,.
\end{equation}
The typical length-scales and masses of possible solitonic configurations can therefore be measured in units of $(e F_\pi)^{-1}$ ($F_\pi e^{-1}$ respectively).

\subsection{The appearance of solitonic configurations (skyrmions)}

In the given setup one can find spherically-symmetric solitonic configurations (``skyrmions"). These are  non-trivial static configurations of $U$, which are regular in the whole space and which minimize the energy functional,
\begin{equation}
E = \int d^3x \mathcal{H}_S\,.  
\end{equation}
Here $\mathcal{H}_S$ is the skyrme Hamiltonian density corresponding to the skyrme Lagrangian $\mathcal{L}_S$. These spherically-symmetric regular static configurations can be found by using a hedgehog ansatz, 
\begin{equation}
\frac{\pi_a}{F_\pi}= F(r) n_a \, ,
\label{hedgehog}
\end{equation}
where $n_a$ is a unit vector in radial direction,  and by minimizing $E$ using the boundary conditions,
\begin{equation}
F(0)= B \pi, F(\infty)= 0 \,,
\label{boundarycond}
\end{equation}
for the profile-function $F(r)$ \cite{adkins1, adkins2}. Here $B$ is a natural number, which can be interpreted as the topological charge of the soliton \cite{adkins1},
\begin{equation}
B = \int d^3x J_0 \, ,
\label{Bcharge1}
\end{equation}
where $J_0$ is the zeroth component of the skyrm topological Chern-Simons current
\begin{equation}
J_\mu = -\frac{\epsilon_{\mu \nu \alpha \beta}}{24 \pi^2} \mathrm{Tr} \left(U^{-1}\partial^\nu U U^{-1}\partial^\alpha U U^{-1} \partial^\beta U\right)\,.
\label{current}
\end{equation}

In the case of $m_\pi = 0$ the solution profile-function scales for large $r$ as \cite{adkins1}
\begin{equation}
F(r) \sim \frac{1}{r^2} \,, 
\end{equation}
whereas for a finite pion mass $m_\pi$, it scales for large $r$ as \cite{adkins2}
\begin{equation}
F(r) \sim \frac{e^{-m_\pi r}}{r} \, .
\end{equation}
For the existence of these solitonic configurations the term $\mathcal{L}_4$ in the skyrme Lagrangian is crucial. Without the term $\mathcal{L}_4$ no non-trivial static lowest energy configuration can exist due to Derrick's theorem \cite{derrick}.

\subsection{Gravitating skyrmions and skyrmions as black hole hair}

We shall now review how and in what sense the above-discussed  skyrmions appear as classical ``hair" of a black hole in the case of the skyrmion black hole solutions discussed in \cite{luckock, droz1, bizon, shiiki, droz2} \footnote{In \cite{luckock, droz1, bizon, shiiki, droz2} the solutions are discussed for the case of zero pion mass $m_\pi$. We extend these solutions to cases with a finite pion mass.}. We focus on cases which are spherical-symmetric and which carry a skyrmion hair, which corresponds to a skyrmion in flat space-time with $B=1$.\footnote{Spherical symmetric skyrmion black holes for $B > 1$ do not exist, since they decay due to one particle decay \cite{glend}. Non spherical-symmetric skyrmion black holes with $B > 1$ are only known for the case which corresponds to the flat space-time skyrmion with $B = 2$ \cite{shiiki2}. By \textit{skyrmion black holes corresponding to the skyrmions in flat space-time with topological charge $B$} we mean skyrmion black holes for which in the limit of vanishing event horizon size skyrmions with topological charge $B$ emerge.} In the appendix we point out how this can be generalized to the case of a gas of many $B=1$ skyrmions.

We couple the skyrme Lagrangian (\ref{lagrangian}) to gravity by using the corresponding energy momentum tensor
\begin{equation}
T_{\mu \nu}^S = \frac{2}{\sqrt{-g}}\frac{\delta \left(\sqrt{-g} \mathcal{L}_S\right)}{\delta g^{\mu \nu}} \,,
\label{energymomentum}
\end{equation}
as a source. As an ansatz for the metric $g_{\mu \nu}$ we use
\begin{equation}
ds^2 = N^2(r)\left(1-\frac{2M(r)G_N}{r}\right)dt^2 - \left(1-\frac{2M(r)G_N}{r}\right)^{-1}dr^2 - r^2 d\Omega^2 \,.
\label{metricansatz}
\end{equation}
In the following we use the dimensionless quantities $x = e F_\pi r$ and $m(x) = e F_\pi G_N M(r)$.

Using the hedgehog ansatz (\ref{hedgehog}) for $\pi_a$ there are in total the three unknown functions - $M(r)$, $N(r)$ and $F(r)$ - for which the Einstein equations,
\begin{equation}
G_{\mu \nu} = 8 \pi G_N T_{\mu \nu}^S \,,
\label{einequ}
\end{equation}
can be solved. (In the appendix we provide the non-vanishing components of the Einstein equations explicitly). For the boundary conditions $N(\infty)= 1$ and $F(\infty) = 0$ non-trivial regular solutions of the Einstein equations exist for several choices of the other two boundary conditions (one boundary condition for $m(x)$ and a second boundary condition for $F(x)$). For the case of $m_\pi = 0$, the solutions have been discussed for example in \cite{droz1} and \cite{bizon}. There exist two different classes of solutions. The classes are characterized by the existence (non-existence resp.) of a value $x_h$ such that
\begin{equation}
1 - \frac{2m(x_h)}{x_h} = 0 \,.
\end{equation}
The solutions in the class where such a value of $x_h$ does not exist are often called ``gravitating skyrmions". The solutions of the other class (indeed the solutions with event horizon) are often called ``skyrmion black holes". They are (to our knowledge) the only known examples of classical black hole solutions where the presence of a skyrmion manifests itself in terms of a hair in the sense that it changes the geometry of the spacetime outside of the black hole event horizon when compared to a Schwarzschild black hole of the same ADM mass.

Given the boundary conditions $N(\infty) = 1$ and $F(\infty) = 0$, all known non-trivial regular solutions within these two classes can be uniquely characterized by $x_h$ and the two dimensionless parameters \footnote{To be precise, there is a degeneracy in the case $x_h \neq 0$: If for a given combination of $\alpha$, $\beta$ and $x_h \neq 0$ a solution to the Einstein equations exists, there also exists a second solution for the same combination. Indeed, for $x_h \neq 0$ there exist two branches of solutions corresponding to two different possible choices of the second boundary condition for $F(x)$ (which we chose at $x = x_h$ when solving the boundary value problem numerically). \cite{bizon} (For $x_h \rightarrow 0$ these two possible boundary conditions $F(x_h)$ for which solutions of the Einstein equations have been found both continuously approach $F(x_h = 0) = \pi$.  For $x_h \rightarrow 0$ the two branches of solutions coincide.) In this work we however only consider solutions from the upper branch which are stable against spherically symmetric linear perturbations, whereas the solutions from the lower branch are not \cite{droz2}.}
\begin{equation}
\alpha = 4 \pi G_N F_\pi^2 \, ,
\label{alpha}
\end{equation}
\begin{equation}
\beta = \frac{m_\pi}{\hbar e F_\pi} \,.
\end{equation}

\begin{figure}
\includegraphics[scale=0.3]{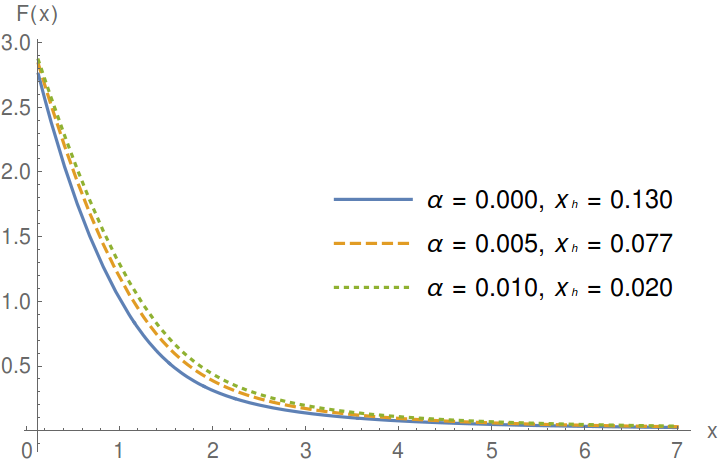}
\caption{Profile function $F(x)$ for the case $\beta = 0$ and $m_{ADM} = 0.065$}
\label{fig:1}
\end{figure}

There exist, however, only certain combinations of values of $\alpha$, $\beta$ and $x_h$ for which non-trivial regular solutions have been found. In order to understand this parameter-space of solutions from a physical point of view, it is useful to elaborate on the physical meanings of $\alpha$ and $\beta$. First, by definition, $\alpha$ and $\beta$ set the ratios of the relevant length scales in the system: $\alpha$ scales as the ratio of the characteristic length scale of the skyrmion $L$ and the gravitational radius of the skyrmion $L_g \sim 2 M_S G_N$
\begin{equation}
\alpha \sim \frac{L_g}{L} \,,
\label{alphascaleradius}
\end{equation}
and $\beta$ is defined as the ratio of $L$ and the Compton wavelength of one pion $L_C$
\begin{equation}
\beta = \frac{L}{L_C} \, .
\end{equation}
Second - due to Witten \cite{witten1} - we know that $\alpha$ scales linearly with the number of colors $N_C$:
\begin{equation}
\alpha \sim N_C \, . 
\label{alphascale}
\end{equation}
Since $L$ and $L_C$ are independent of $N_C$, $\beta$ does not scale with $N_C$. We therefore fix $\beta$ and review the parameter space of solutions of the Einstein equations with boundary conditions $N(\infty) = 1$ and $F(\infty) = 0$ for fixed $\beta$ \cite{bizon}:

There exists a maximal value $\alpha_{max}$ and non-trivial solutions have only been found for $0 \leq \alpha \leq \alpha_{max}$ (in the limit $\alpha \rightarrow 0$, which can be interpreted either as the decoupling limit $G_N \rightarrow 0$ or as the limit $N_C \rightarrow 0$, the solutions go over to Schwarzschild solutions). According to (\ref{alphascaleradius}) and (\ref{alphascale}), the maximal value of $\alpha$, for which non-trivial solutions of the Einstein equations are known,  corresponds both to a maximal value for the ratio $\frac{L_g}{L}$ and to a maximal value of the number of colors $N_C$ for which such non-trivial solutions are known. From the physics point of view,  this bound on  $\alpha$ means on the one hand that non-trivial solutions only exist as long as the skyrmion size $L$ is bigger than its own gravitational radius $L_g$.  In other words, non-trivial solutions exist as long as the skyrmion itself is not a black hole. On the other hand, the existence of $\alpha_{max}$ means that non-trivial solutions only exist as long as the number of colors is bounded from above by a maximal value $N_*$. We shall show below that, when taking into the account that $L$ is set by the QCD length, these two physical meanings become unified. We shall show that one can explain the appearance of the maximal value $\alpha_{max}$ as a particular manifestation of the so-called general black hole bound on the maximal allowed number of particle species in any given theory \cite{giabound, *giabound2, *giabound3}.

On top of that, for each given value of $\alpha$ in the allowed range there exists a maximal possible event horizon size $r_h^{max, \alpha, \beta}$ and for each given value of $\alpha$ in the allowed range skyrmion black holes have only been found in the range $0 \leq x_h \leq x_h^{max, \alpha, \beta}$ (in the limit $x_h \rightarrow 0$ the solutions go over to the solutions of gravitating skyrmions). $x_h^{max, \alpha, \beta}$goes to zero for $\alpha \rightarrow \alpha_{max}$ , in particular $x_h^{max, \alpha_{max}, \beta} = 0$. Roughly speaking, the presence of a maximal possible value $x_h^{max, \alpha, \beta}$ means that the sum of the horizon size $r_h$ and the gravitational radius of the skyrmion $L_g$ cannot become larger than the typical skyrmion size $L$. Indeed, it follows from the existence of $x_h^{max, \alpha, \beta}$ that for each given value of $\alpha$ within the allowed range a relation $r_h^{max,\alpha,\beta} + L_g \sim L$ holds. The skyrmion black holes are therefore configurations where an event horizon of size $r_h$ is formed \textit{inside} the skyrmion.  The skyrmion then represents a ``hair" in the sense that it is not fully swallowed by the black hole.
\newline


All solutions for $F(x)$ scale for large $x$ as in the flat space-time case, i.e.,  as $\frac{1}{x^2}$ for $\beta = 0$ and as $\frac{e^{-m_\pi x}}{x}$ for $\beta \neq 0$. The solutions for $m(x)$ go to a constant value $m_{ADM}$\footnote{$m_{ADM}$ is related to the ADM mass $M_{ADM}$ by $m_{ADM} = e F_\pi G_N M_{ADM}$} for large $x$. For small $x$, however, $m(x)$ significantly differs from this constant value. Therefore, the presence of the skyrmion hair changes the 
near-horizon geometry of the space-time when compared to a Schwarzschild black hole with the same ADM mass.

 For given $x_h$, $\alpha$ and $\beta$ within the allowed range the solution functions $F(x)$, $m(x)$ and $N(x)$ are only known in the regime $x \geq x_h$ (outside of the event horizon).\footnote{See however \cite{tamaki} for some studies on the internal structure.} The values $m(x_h)$, $F(x_h)$ and $N(x_h)$ are known explicitly for each solution. 
For more details about the solutions and the numerical methods used for 
obtaining them the reader is referred to \cite{luckock, droz1, bizon, shiiki, droz2}.

\begin{figure}
\includegraphics[scale=0.3]{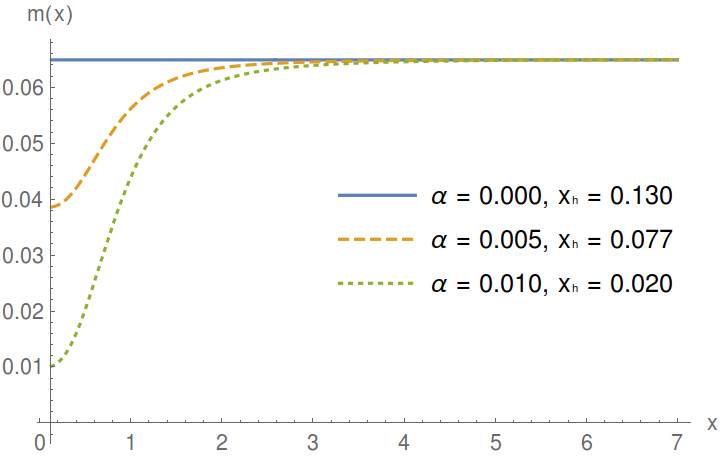}
\caption{$m(x)$ for the case $\beta = 0$ and $m_{ADM} = 0.065$}
\label{fig:2}
\end{figure}

\begin{figure}
\includegraphics[scale=0.3]{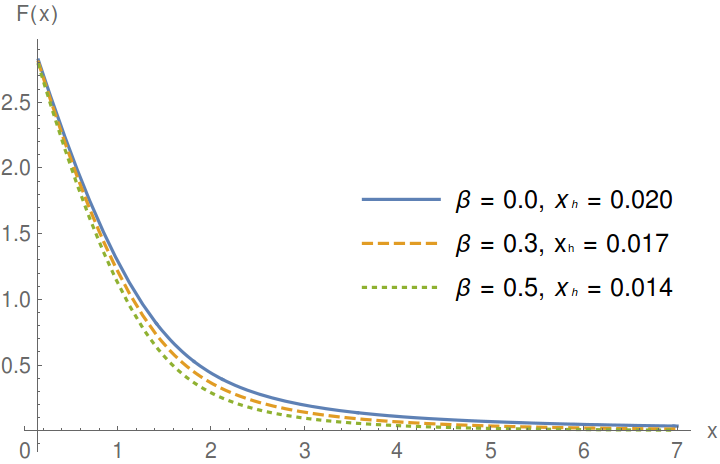}
\caption{Profile function $F(x)$ for the case $\alpha = 0.01$ and $m_{ADM} = 0.065$}
\label{fig:3}
\end{figure}

\begin{figure}
\includegraphics[scale=0.3]{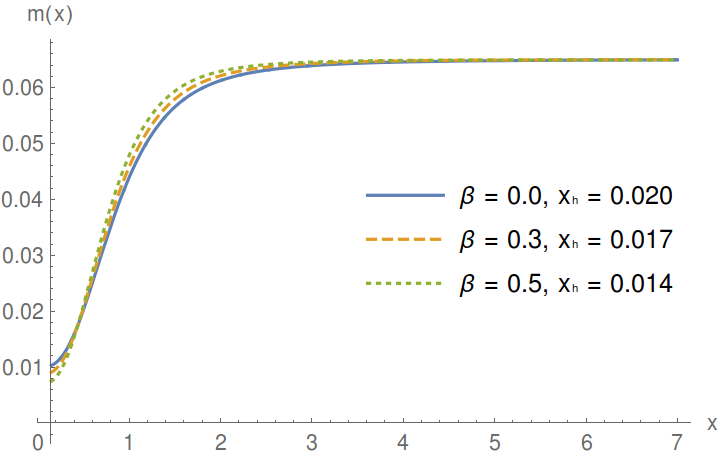}
\caption{$m(x)$ for the case $\alpha = 0.01$ and $m_{ADM} = 0.065$}
\label{fig:4}
\end{figure}

\begin{figure}
\includegraphics[scale=0.3]{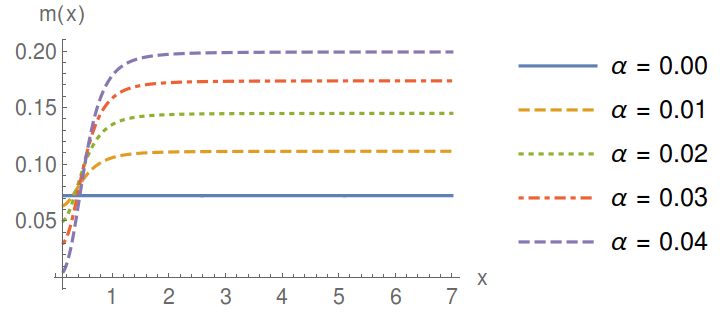}
\caption{$m(x)$ for the case $\beta = 0$ and $x_h = x_h^{max,\alpha, \beta = 0}$}
\label{fig:5}
\end{figure}

For illustration we plot some numerical solutions for $F(x)$ and $m(x)$ for different cases of $\alpha$, $\beta$ and $x_h$. In Figure \ref{fig:1} and Figure \ref{fig:2} we fix $\beta = 0$ and $m_{ADM} = 0.065$ and plot $F(x)$ ($m(x)$ respectively) for $\alpha = 0$, $\alpha = 0.005$ and $\alpha = 0.01$. In Figure \ref{fig:3} and Figure \ref{fig:4} we fix $\alpha = 0.01$ and $m_{ADM} = 0.065$ and plot $F(x)$ ($m(x)$ respectively) for $\beta = 0$, $\beta = 0.3$ and $\beta = 0.5$. In Figure \ref{fig:5}
we fix $\beta = 0$ and plot $m(x)$ for $x_h^{max,\alpha = 0, \beta = 0}$, $x_h^{max,\alpha = 0.01, \beta = 0}$, $x_h^{max,\alpha = 0.02, \beta = 0}$, $x_h^{max,\alpha = 0.03, \beta = 0}$ and $x_h^{max,\alpha = 0.04, \beta = 0}$.

\section{Skyrmion black holes as baryonic black holes and conservation of baryon number in systems with black holes}

We argue that the skyrmion black holes we discussed are natural candidates which can be used in an argument suggesting that there is no conflict between the conservation of baryon number and semi-classical black hole physics.
For this purpose we first point out in subsection A that in the simplified case when we restrict our analysis to pions, the skyrmion black holes (which - as reviewed in section II - in particular have an event horizon which is located inside the skyrmion) are nothing but baryonic black holes with event horizon located inside the baryon. 
In section B we review the standard folk theorem and carefully analyze the assumptions on which 
its proof is based.  This analysis allows us to prepare a ground for understanding how the skyrmion black holes provide a loophole 
in this proof by showing the inconsistency of some of its  key assumptions.  
In subsection C we then provide a Gedankenexperiment,  which explicitly shows how the existence of black holes with skyrmion hair abolishes the folk theorem argument for baryon number non-conservation in systems with black holes. 
  In subsection D we discuss how a modified version of the same  Gedankenexperiment provides a counter-argument to a so-called weak gravity conjecture \cite{weak1, *weak2}. 

In this section we work in the limit of massless quarks, indeed we restrict our analysis to $\beta = 0$.

\subsection{Mapping between skyrmion black holes and baryonic black holes}

 In this section we shall provide a mapping between skyrmion and baryonic black holes.  
 For this purpose we shall employ the two observers (Alice and Bob) describing strong interactions in terms of high and low energy degrees of freedom
  respectively.    
From the point of view of Bob the elementary degrees of freedom of the strong interaction are pions. Only a high-energy observer, Alice, can resolve the quark substructure of pions. For Alice QCD is the appropriate theory for  describing the strong interactions. Bob can describe strong interactions in a low-energy effective theory of QCD in terms of chiral Lagrangians.  The pions which from his point of view are the fundamental degrees of freedom are Nambu-Goldstone bosons, which appear since chiral symmetry of QCD is broken spontaneously by the quark condensate.

 Up to four-derivatives, the effective Lagrangian for Bob is described by the Skyrme Lagrangian $\mathcal{L}_S$, which we considered in section II 
  \footnote{Since, in this section we restrict our analysis to the case $\beta = 0$, the term $\mathcal{L}_m$, which we introduced in the Skyrme Lagrangian in section II and which breaks chiral symmetry explicitly, is now absent. For $\beta = 0$ which we consider in this section $\mathcal{L}_S$ is therefore invariant under chiral symmetry.}. Skyrmions therefore appear as solitons in Bob's low-energy effective description of strong interactions. It was argued in \cite{bala, wit1, *wit2} that these skyrmions in the low-energy effective theory correspond to baryons in the high-energy theory (QCD). In other words: What Alice calls ``baryons" are skyrmions from the point of view of Bob. This is because there is a one-to-one matching between the baryon number current in Alice's QCD,
\begin{equation}
J_\mu = \frac{1}{N_C}\sum_f \bar{q}_f \gamma_\mu q_f \,, 
\label{Bcurrent}
\end{equation}
and the topological Chern-Simons current (\ref{current}) in Bob's effective theory,
\begin{equation}
J_\mu = -\frac{\epsilon_{\mu \nu \alpha \beta}}{24 \pi^2} \mathrm{Tr}\left(U^{-1}\partial^\nu U U^{-1}\partial^\alpha U U^{-1}\partial^\beta U\right) \, .
\label{Tcurrent} 
\end{equation}
Thus, the black holes, which from the point of view of Bob carry skyrmion ``hair", do carry the baryon ``hair" from the point of view of Alice.

\subsection{Conservation of baryon number and folk theorems}

 In this section we shall review the folk theorem that suggests the necessity of baryon number violation by black holes and 
 prepare the basis for understanding why the existence of black holes with skyrmion hair abolishes this conclusion.    
 
 The standard folk theorems argument is very simple and elegant. 
Consider an {\it arbitrarily-large}  black hole in which a certain number of particles with some global charge 
is thrown.  We shall refer to this charge as  {\it baryon number}. 
 Assuming that the black hole carries no hair under such global quantum numbers, the resulting 
 classical metric is described by the Schwarzschild
(or Kerr) metric irrespective of the initial baryonic charge that went into the black hole.  The black hole then slowly evaporates as thermal up until it reaches a size comparable to the quantum gravity scale, e.g., the Planck length, $L_P$. Until this stage, 
due to the thermality of the outgoing radiation, no (or negligible) net baryon charge is emitted.  
 Thus, the  black hole retains almost its entire initial baryon number during the evolution process.  But, after the black hole reaches the Planck size, it is too late to 
return back the initial charge, which can be {\it arbitrarily big}.
  Hence, one concludes that the baryonic charge cannot be conserved in the presence of a black hole\footnote{ An alternative would be to postulate an infinite number of degenerate remnant species crowded around the Planck mass, but this possibility creates standard unitarity problems, and shall not be considered here, especially since we are offering a much less costly way out.}.

  In order to understand where the above argument goes wrong, let us examine the picture carefully. 
 
 This argument heavily relies on the two assumptions:
 
 {\it 1) Non-existence of any type of (semi)classically-measurable hair 
 that could carry information about the black hole baryonic charge; 
 
 and 
 
 2)  Exact thermality of the emitted Hawking radiation. }  \\
   
 In fact, these  two assumptions are not unrelated.  In particular, the  inconsistency of the first assumption, would most probably also imply  the inconsistency of the second one:  if a classical baryonic hair existed, it would presumably affect the black hole metric and  cause some correction to the emission process.  For example, it would create an asymmetry between the emission rates of baryons and anti-baryons. 
 The converse is not necessarily true: large deviations from thermality  would not necessarily imply the existence of a classical baryonic hair that could guarantee the recovery of the initial baryon number.  
 
  The main purpose of the present work is to argue against the first assumption, as this is the 
 key for the semi-classical recovery of the baryonic charge.  
 Nevertheless, it is instructive to understand the role of both assumptions. 
      
 Let us discuss the deviations from thermality first.  The standard picture ignores the role of deviations from the thermal emission, 
 implicitly assuming that  such deviations 
 are exponentially-small up until the black hole size diminishes to  Planck length. If so, throughout this period 
only a negligible (exponentially-small) net baryon charge can be emitted.   
   Notice,  that the value of the baryon mass plays no essential role in this suppression.  In order to see this, let  us split the black hole evaporation epoch in two periods, defined by the black hole temperature ($T$) being below or above the baryon mass ($m_B$).
   We shall assume that the initial black hole is big enough, with the event horizon size $r_h \gg  {\hbar \over m_B}$, so that 
   the temperature is much lower than the baryon mass, $T \ll m_B$.   We shall treat $m_B$ as a free parameter, which 
   can be arbitrarily small or large.   
 During the low-temperature period, $ m_B \gg T$,  the rate of baryon emission is exponentially suppressed by the Boltzmann factor, $\Gamma_B \, \sim \, e^{-{m_B\over T}}$. Correspondingly, the baryon number produced by Hawking radiation in this epoch is exponentially-small. 
 
 However, this continues to be true  even after the temperature exceeds the baryon mass:  production of baryon number is exponentially suppressed as long as the emission keeps being thermal. The reason is that, although the baryon emission rate is no longer exponentially-suppressed, it is exactly equal 
 to the analogous emission rate of anti-baryons, $\Gamma_{B} = \Gamma_{\bar{B}}$.  As a result of this equality, no net baryon number is produced.  Thus, the departure from the thermal equilibrium 
 is a necessary condition for producing a net baryon charge in black hole evaporation.  This condition is in a way analogous  to 
 Sakharov's necessary condition for generating the net baryon number in the early Universe. In both cases out-of-equilibrium departure is required.

  We must note here that a black hole itself has a built-in source of departure from exact thermal equilibrium, due to its back reaction on the temperature \cite{thermalitydeparture}. The exact thermality cannot hold, due to the very fact that the temperature must change during the evaporation process. 
 It is impossible to keep evaporation exactly thermal while changing the temperature with time.
 The time-derivative of temperature measured against the temperature-squared, ${\dot{T} \over T^2}$, serves as a measure of departure from thermal equilibrium \cite{speciesstrings}. 
  For a black hole this parameter is equal to ${\dot{T} \over T^2} =  \left ({T \over M_P }\right )^2$,   
 which shows that deviations from thermality are not exponentially-small \cite{thermalitydeparture}.  
  However, this correction alone, despite being much stronger than what is usually assumed,  is still tiny and not nearly-enough
 for producing a net baryon number of required magnitude. 
  So we need a stronger source for the recovery of the baryonic charge. 
  
   As we shall argue in subsection C,  such a source is provided by 
  the classical skyrmion hair.  It is this hair that carries the information about the black hole baryonic charge and makes  its conservation possible. 
     The existence of this classical baryonic/skyrmion hair renders both of the above assumptions of the standard  folk theorem inconsistent.  Indeed, we shall  argue in subsection C that the assumption that baryon number is conserved in systems with black holes does not necessarily lead to a contradiction with black hole physics.

   As we shall see, because the skyrmion-type hair carries the information about the black hole baryon charge,  the new characteristic length scale is created in form of the scale $L$.  After being shrank to this size (or equivalently after reaching 
   the temperature $T  =    {\hbar \over L}$) the black hole undergoes a transition 
  revealing its baryonic charge in form of the classical skyrmion hair.    Unlike $L_P$,  the skyrmion scale $L$ is not a fixed UV-scale of gravity, but rather is a characteristic of a given black hole and depends on its baryonic charge.   
 Thus,  the critical size of a black hole beyond which both the no-hair conjecture as well as Hawking's thermal approximation break down is determined by the existence of the baryonic charge of a black hole. This fact  consistently accommodates the charge-conservation.

\subsection{Conservation of baryon number in systems with black holes}

 Using the mapping between the baryon number current and skyrm topological current,  
 we shall now describe a Gedankenexperiment  in which conservation of the baryonic charge by a black hole is consistently accommodated.  

This Gedankenexperiment is identical to the one used in the folk theorem, but it takes into the account 
 the existence of skyrmion black holes as well as the matching  between the skyrmion and baryon currents. It is this new information that invalidates the old conclusion.
 We shall discuss three versions of the Gedankenexperiment in which 
 the emergence of the baryonic charge in form of a skyrmion hair ranges from 
 being a logical possibility all the way to being a necessity.

\subsubsection{The setup}

  For simplicity, we shall illustrate the point with a single skyrmion for which the solution  is well-understood
 in various regimes.  We shall assume the framework of $SU(N_C)$ QCD with $N_C$-colors and  massless 
 quark flavors  and take the 't Hooft's
 large-$N_C$ limit, keeping the QCD scale fixed.  The QCD length then sets the size of the skyrmion/baryon, $L$, whereas the mass 
 of the skyrmion/baryon  scales as $M_S \sim N_C {\hbar \over L}$.       
 Thus, using $L$ and $N_C$ as two independent control parameters we can design a skyrmion/baryon of different possible masses and  sizes.  In a theory with zero gravity ($G_N = L_P=0 $) these parameters can be arbitrary.  However,  as we discuss in the Appendix C, with gravity taken into the account an important bound appears, 
 \begin{equation}  
     N_c < N_* \equiv \left ({L \over L_P} \right)^2 \, .      
 \label{species} 
 \end{equation}   
 In order to understand the physical significance of this bound notice that using the definition of $N_*$ given in (\ref{species})  
 the relation between the gravitational radius of baryon ($L_g$) and its non-gravitational size ($L$) set by the QCD length can be recast as, 
 \begin{equation}
     \alpha \sim {L_g  \over L} = {N_C \over N_*} \, .
     \label{ratioL}
  \end{equation}      
 This equation tells us that for $N_C > N_*$ the gravitational radius of baryon exceeds the QCD length, L, 
 i.e., the physical size of baryon set by QCD.  As we show in the appendix C, in this regime the same is true  about any state with
 $N_C$-iality: all such states are localized within their own gravitational radii, i.e., they are all black holes.   
  In other words, 
 there exist no baryons in such a theory! 

 As we discuss in the Appendix C, the above phenomenon has a deeper physical meaning.  Remarkably, it reveals a ``secret"  information about the validity domain of QCD theory, when the latter is coupled to gravity.   
 Namely, the above bound coincides  with the black hole bound on the number of particle species in a generic theory, derived in \cite{giabound, giabound2, giabound3}.  When applied to the current situation (see Appendix C) the latter bound is telling us 
that for $N_C >N_*$, the length scale at which the quantum gravity effects become strong exceeds the QCD length, $L$,  i.e., becomes larger than the size of a 
baryon or a glueball.  The significance of this result is that  for $N_C >N_*$  it becomes impossible to treat QCD as a well-defined theory. 
Correspondingly,  the question of violation of baryon number becomes meaningless, since there exist no baryons within the semi-classical theory of gravity.  The reader is referred to the Appendix C for this discussion.   
 Thus, we shall restrict our analysis by (\ref{species}).  In particular, for definiteness we shall take $N_C \ll N_*$, and thus 
 $L_g \ll L$.

Now, consider the following setup.  Let us take a  black hole with event horizon size, $r_h$, much bigger than  the QCD length 
$L$, i.e., much bigger that the baryon/skyrmion size. 
  Let us assume that a single baryon is thrown into this black hole.   Of course, by construction the initial mass 
of the black hole is much larger than the baryon mass. We shall assume that the black hole is electrically neutral and 
has negligible angular momentum.   Such a black hole then emits the Hawking radiation and becomes smaller due to evaporation.
  In order to re-emphasize,  the quantities $L$ and $N_C$ (and thus, $L_g$) are the fixed parameters of the theory. 
 The quantity that changes during the thought experiment is the event horizon of a black hole $r_h$. 
 
\subsubsection{The Gedankenexperiment $I$}

  We shall now follow the evolution of the above black hole  and show that - contrary to the folk theorem's conclusion -   no logical inconsistency with conservation of the baryon number arises.

We shall distinguish the following two regimes.  The first one is the regime in which the event horizon of a black hole is much larger than the size of the baryon/skyrmion, $r_h \gg L$. In this regime there exist no known
black hole solutions with classical baryon/skyrmion hair.
 Notice, this fact {\it per se} cannot serve as an argument in favor of non-conservation of baryonic charge, since there is a big difference between inaccessibility of a given quantum number in classical measurements and its violation.   Moreover, as we shall discuss below, the topological skyrmion charge 
 is encoded in a  surface integral. 

 Hence, when a baryon/skyrmion is thrown into a black hole with $r_h \gg L$ the data about the baryonic charge can 
 continue to exist, even though it may not be accessible classically for an outside observer as long as the black hole - which evaporates due to the emission of Hawking radiation - remains in the regime $r_h \gg L$.  The crucial difference  in case of the standard ``folk theorem" reasoning is that there 
no analog of the scale $L$ exists, or to put is better, it is equal to the Planck length, since this is the only scale at which a departure from classical no-hair assumption is assumed to be possible.  This is what makes - under that reasoning -  
the baryon number conservation logically inconsistent. 

 This is not true in the present case, because the existence of the scale $L$ creates a second regime in which the event horizon of a black hole is smaller than the size of the baryon/skyrmion, $r_h < L$. What makes the
crucial difference between the two regimes considered is that in the second regime classical black hole solutions with skyrmion hair do exist and this creates a logical possibility of re-emergence of the baryonic charge in form of the 
skyrmion hair. 

  We are unable to identify any logical inconsistency with such a scenario, and correspondingly  we do not see any evidence, at least in this setup, of incompatibility of conserved baryon charge with semi-classical black hole physics.

\subsubsection{The Gedankenexperiment $II$: Monitoring Baryon/Skyrmion Hair by  a Topological Surface Integral}   

We shall now try to strengthen the argument in favor of emergence of the classical skyrmion hair by endowing the asymptotic observers with an ability to measure a certain surface integral that carries information about the skyrmion charge.

   Let us first notice that the skyrmion topological current  (\ref{current})  represents a 
   Hodge-dual of the exterior derivative of the following two-form
   \begin{equation}
   S_{\mu\nu} \equiv  - {1\over 4\pi^2} \left ( F(r) - {1\over 2} {\rm sin}(2F(r)) - \pi \right ) 
   \partial_{[\mu} {\rm cos} (\theta) \partial_{\nu]} \phi    \,.  
   \label{twoform}  
  \end{equation} 
   Here $r, \theta$ and $\phi$ are the usual spherical coordinates and $F(r)$ is 
   the  profile function, with the boundary conditions $F(0) =  \pi$ and 
   $F(\infty) = 0$.  (Here we have chosen the $B=1$ case of (\ref{boundarycond}).) 
    Notice, that the choice of the constant $\pi$ in (\ref{twoform}) is uniquely dictated by 
    the requirement that the two-form  $S_{\mu\nu}$ is well-defined 
    everywhere.  This is exactly how the information about the non-trivial 
    topology  is carried by this two-form.

 Since this two-form is smooth everywhere, one can use Stokes theorem with a boundary surface (for example a two-sphere $S_2$) at infinity to express the topological charge (\ref{Bcharge1}) as an integral over that boundary surface,
   \begin{equation} 
     B = \int_{S_2} dX^{\mu}\wedge dX^{\nu} S_{\mu\nu}   =  1\, ,
   \label{Intboundary} 
   \end{equation}  
where $X^{\mu}$ are the world-volume coordinates of the boundary sphere. Note, that the choice of the constant $\pi$ in (\ref{twoform}) is crucial at this point because only for this choice the two-form is smooth everywhere and Stokes theorem can be consistently applied with (only) choosing a two-sphere at infinity as boundary surface. This allows to consistently define the topological charge (\ref{Bcharge1}) on a two-sphere at infinity. Once the two-form (\ref{twoform}) is defined with the correct constant, this definition must be used at any time, in particular even after the skyrmion/baryon is swallowed by a black hole.
    
  We can now repeat the same Gedankenexperiment as in the previous section and insert a skyrmion/baryon into a black hole.   The asymptotic observers Alice and Bob now can monitor the skyrmion charge if they can measure the above surface integral. 
   The possibility of such a measurement  is not at all a trivial assumption.  First,  it requires the
   existence of a string-like object that is coupled to the two-form $S_{\mu\nu}$ via the usual coupling between the string world-sheet element and an antisymmetric two-form,
   \begin{equation} 
     {\mathcal S}_{coupling}  = g \int  dX^{\mu}\wedge dX^{\nu} S_{\mu\nu} \, ,  
   \label{couplingtostring} 
   \end{equation}  
   where $X^{\mu}$ are string embedding coordinates and $g$ is a coupling constant.
Second, the coupling constant $g$ between the string and the two-form must be fractional 
   (in units of skyrme topological charge).
If these two assumptions are satisfied, an interference experiment which leads to a non-trivial phase shift, similarly to the Aharonov-Bohm effect, can be performed by fully enclosing the boundary two sphere around which the integral (\ref{Intboundary})  is taken with the two-dimensional world-volume of a string loop. The change in the action obtained through such a (stringy generalization) of the Aharonov-Bohm effect is then given by
\begin{equation}
\Delta \mathcal{S} = 2 \pi g
\end{equation}
The corresponding Aharonov-Bohm phase shift can be measured if $g$ is fractional. In this sense the two-form $S_{\mu\nu}$ becomes physical when $g$ is fractional. Correspondingly, if $g$ is not fractional, then the two-form is unphysical.\footnote{One may wonder how the physicality of a two-form can depend on the charges of other (spectator) objects in the theory. Such situations are however well-known. For example, this is similar to the question of physicality of the Dirac string, which depends on the existence of fractionally charged particles (in units of magnetic charge which is the same as the magnetic flux carried by the Dirac string). If a fractionally charged particle existed, the Dirac string would be physical as it can be detected experimentally by an Aharonov-Bohm type phase shift in the experiment in which a particle is transported along a closed path encircling the Dirac string. Dirac needed to avoid exactly this: Dirac demanded his string to be unphysical and this automatically lead to the requirement of Dirac electric charge quantization (in units of inverse magnetic charge). Our situation is identical: If we demand that the charge of the string $g$ is quantized in units of inverse skyrmion charge, the two-form 
$S_{\mu\nu}$ will become unphysical, since the only way to detect it is through (stringy generalization of) Aharonov-Bohm experiments in which the string world-volume encloses the skyrmion at infinity. On the other hand, if the string charge is fractional, the two-form becomes physical and correspondingly the presence of skyrmion can be detected at infinity. At the level of the effective field theory discussion there is no inconsistency in existence of probe strings of arbitrary charge and hence 
no obstacle in making skyrmion observable via Aharonov-Bohm phase shift.}
    We shall assume that Alice and Bob have such spectator strings with $g$ fractional at their disposal and can therefore monitore the skyrme charge at infinity by measuring the Aharonov-Bohm phase shift.

     It does not really matter, how we manage to insert the skyrmion/baryon inside the black hole with $r_h \gg L$.  For example, we can throw it in a pre-existing black hole or alternatively we can collapse a skyrmion together with some other non-baryonic matter into the black hole. 
   As long the initial topological charge measured by the surface integral 
 (\ref{Intboundary}) is non-zero, the observer has a knowledge about the baryon charge that went into a black hole.  The power of topology is that 
 it is enough to know that $F(0) =\pi$ at the initial stage when the baryon/skyrmion is not  inside of the black hole.  This fixes the topological charge.   Of course, after the baryon/skyrmion is swallowed by the black hole,  
 an external observer can no longer access the point $r=0$, but this is irrelevant, since the information about the charge is encoded in the boundary surface 
 integral given by (\ref{Intboundary}).

    What happens once the event horizon $r_h$ of the black hole
   becomes - due to Hawking evaporation - of order (or smaller)    
 the QCD length $L$, which sets the typical size of a skyrmion/baryon?  

  Let us first eliminate the possibility that the black hole can shrink further down to the size 
  $r_h \ll L$, without leaving a classical skyrmion hair behind.\footnote{Note, that this will show that, when a black hole shrinks to a size L, a classical skyrmion hair has to be left behind. We want to emphasize that this does not necessarily mean that this skyrmion hair has to be of the same kind as the skyrmion hair discussed in section II. The skyrmion hair discussed there is a natural candidate for such a hair, we can however not exclude the option that (so far unknown) classical skyrmion hair of different kind (not covered by the black hole solutions discussed in section II) exist and that the hair of a realistic black hole which shrinks to a size $r_h \ll L$ is correctly characterized by such a - yet unknown - skyrmion hair.}  We shall eliminate this possibility by contradiction. 
   Hence, let us assume that this is the case, i.e.,  our black hole reached a size  $r_h \ll L$ without revealing a skyrmion hair. 
        Let us draw a sphere of radius $R$ around the black hole. Let us  choose the radius to be in between the two scales,   $r_h \ll \ R\ll L$.   Both Alice and Bob can measure the surface integral (\ref{Intboundary}) and thus detect the skyrmion charge of a black hole. What would they conclude? 

\begin{figure}
\centering
\includegraphics[scale=0.4]{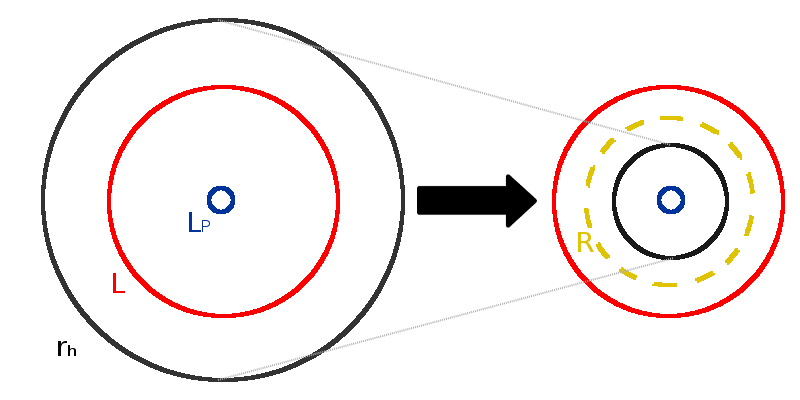}
\caption{Illustration of two stages of the black hole evolution in the case of the first possibility: A black hole with event horizon $r_h$ (black circle) much bigger than the skyrmion/baryon size $L$ (red circle) shrinks to a black hole configuration with event horizon $r_h \ll L$. The blue circle represents the Planck length $L_P \ll L$. The yellow circle represents the sphere of radius $R$ around the black hole used in the argument to eliminate the possibility that a black hole shrinks to a size $r_h \ll L$ without leaving a classical skyrmion hair of some sort behind.}
\label{fig:6}
\end{figure}

  Since, we are at distance $R\ll L$, we are in the effective theory of quarks and gluons, and Alice 
  will attribute the surface integral to the quark current 
  (\ref{Bcurrent}) of QCD.  By the baryon/skyrmion correspondence, this current is mapped on the skyrme topological Chern-Simons current (\ref{Tcurrent}) and thus Bob should attribute the same measurement to this topological charge. 
  But a non-vanishing value of this charge requires the existence of a classical skyrmion configuration, which has a size $L$.\footnote{We assume that gravity is not producing new states with baryon/skyrmion number that are more compact than the QCD scale $L$.  In such a case a classical skyrmion hair  of course need not emerge at the scale $L$, but at some shorter distance.   However, one way or the other the baryonic charge will be transmuted to these new states.   If one pushes the size of these states all the way to the Planck length, we simply arrive to the picture of remnants . This is the possibility that we have dismissed  (due to the fact that it implies the existence of infinite number of species crowded  at the Planck scale).}
   This contradicts to our original assumption that the black hole shrank to the size $r_h \ll L$, without leaving a classical hair behind. 
     
  Thus, we have eliminated the possibility of no skyrmion hair appearance when the black hole passes  the critical size 
  $r_h < L$.  With this fact established,  there are the two possible scenarios.  
 
  The first possibility is that a black hole shrinks further down and continues to evaporate.   
   In such a case, while shrinking, the black hole must leave the classical skyrmion field behind. That is,  the skyrmion re-appears and the black hole shrinks further down to the Planck-size inside the skyrmion (see Figure \ref{fig:6} for an illustration). In this way, at the end the whole skyrmion/baryon gets recycled 
(i.e.,  the decaying black hole leaves the skyrmion/baryon behind).
 As discussed above,  the re-appearance of a classical baryon/skyrmion hair is enforced by the following two facts. 
 First,  the surface integral measured by Bob and Alice must be matched by the corresponding 
 skyrmion/baryon current, once the black hole shrinks beyond the QCD length.  
 Moreover,  there exists an excellent candidate for  accommodating this matching:  
 since we know that a skyrmion black hole with event horizon $r_h$ which is located inside the skyrmion does exist,  this skyrmion black hole is a natural candidate for describing the final stage of the black hole evolution.   
  
 As an alternative option,  one may assume that the black hole of size $L$ stops to evaporate, because it reaches some kind of an extremal state, being stabilized by the skyrmion charge, so that it  does not radiate further.  
   Although we cannot exclude such an option, it  of course would not work as an argument in favor of baryon number violation. 
 The stabilized black hole would simply carry the original baryon number.    
  Moreover, such a stabilization would also require the reappearance of a classical skyrmion hair around the scale $L$.  Indeed, without it, the black hole metric would continue to be described by Schwarzschild, in which case nothing would 
 prevent it from continuing to evaporate further down.    Thus,  one way or the other, a skyrmion hair must manifest itself around the 
 QCD scale.

\subsubsection{The Gedankenexperiment $III$: Monitoring Baryon/Skyrmion Hair by  a Spectator Field }

 In the previous version Alice and Bob were monitoring black hole baryon/skyrmion 
 charge by measuring the surface integral (\ref{Intboundary}).  
  
 We shall now offer yet another version of the experiment, in which case 
 Alice and Bob can monitor the baryon charge by a {\it classical} measurement.  
 For this we need to design an ``apparatus" that shall enable 
 an outside observer to access the baryonic data of a black hole classically also in the regime  $r_h \gg \, L$.  However, 
 an outside observer should be able to make the  measurement  {\it arbitrarily soft},  in order not to disturb the semi-classical evolution as compared    
to the previous experiment.  In other words, the new agent must be so soft with respect to classical black hole 
dynamics that conservation of baryon charge  cannot  be ``blamed" on its presence.

In order to make the baryonic data classically-accessible also in the first regime  (for $r_h \gg \, L$), we can gauge the baryon number symmetry  {\it infinitely weakly} by introducing a $U(1)$ gauge field $B_\mu$ and an infinitesimally-small gauge coupling $g_B$. (Notice, that in the standard folk theorem case, such a weak gauging would run us into an inevitable contradiction and in fact would lead us to the conclusion that gauge coupling {\it cannot be}  arbitrarily-weak (see below). In the present situation this is not the case, since 
the role of such gauging is only to make the information about the skyrmion hair - which for  
$r_h < L$ exists regardless of gauging the symmetry - classically-accessible for an external observer already at an earlier stage $r_h \, \gg \, L$.)
By choosing the gauge coupling infinitesimally-small we can ensure that the resulting baryon-electric field can have no effect neither on the initial black hole metric nor on the Hawking evaporation process as compared to the un-gauged case. In this sense we call the gauge field a ``spectator". The baryonic data is accessible for an outside observer since for any non-zero $g_B$, an asymptotic observer can - no matter how big the black hole is - measure the electric field of $B_\mu$ at infinity due to  Gauss's law. The integrated  electric flux is as usual given by the total conserved charge,
\begin{equation} 
    Flux \,  = \, g_B Q_B \equiv g_B \int d^3x J_0 \, , 
    \label{flux}
   \end{equation}  
  where  $J_0$ can be computed either using (\ref{Bcurrent}) or  (\ref{Tcurrent}). 
  Both Bob and Alice can measure the above electric flux of $B_{\mu}$.  However, they interpret this measurement differently. From the point of view of Alice $B_\mu$ is sourced by the baryon number current (\ref{Bcurrent}) of QCD, whereas,  from the point of view of Bob $B_\mu$ is sourced by the skyrme topological Chern-Simons current (\ref{Tcurrent}). 
 In this way, Alice and Bob can monitor the corresponding conserved charges throughout the evaporation process.

 The analysis of the experiment, once the black hole reaches the size $r_h \sim L$ 
 is essentially identical to the previous section, and we shall refer the reader to it.
 The only difference is that instead of measuring the topological surface 
 integral (\ref{Intboundary}) Alice and Bob can now (in addition) measure the electric flux (\ref{flux}).

  Notice,  the existence of black holes with skyrmion hair is absolutely crucial for our argument. Without this knowledge, we would wrongly conclude that weak gauging of the baryon number is inconsistent (see the next section),  since 
 the only source for affecting the Hawking radiation would be a Reissner-Nordstrom type modification of the metric due to electric field of $B_{\mu}$, which should vanish for $g_B \rightarrow 0$ limit and cannot help in the recovery of the baryon charge.  
 In other words, $B_{\mu}$ plays no direct role in conservation of the baryon number, its existence only allows us to discover the ``reincarnation" of baryonic 
 hair into the skyrmion hair, which otherwise we would have overlooked. 
 
In all three cases the skyrmions manifest themselves at the scale $L$ when the black hole horizon can be arbitrarily bigger than the Planck length. We can therefore conclude that our refined  Gedankenexperiments abolish the standard argument  
of baryon number violation by black holes. 
\newline

 \subsection{Consequences for weak gravity conjecture}  
 
 The thought experiment with a weakly-gauged spectator field $B_{\mu}$ naturally leads us to 
 another important implication of our results.  
 It is interesting that we simultaneously create a loophole in a so-called weak gravity conjecture \cite{weak1, *weak2}, which states that black hole physics imposes a constraint that the gauge coupling cannot be arbitrarily weak.
  The {\it no-global-charge}  folk theorem can be considered as the limiting case of this weak gravity conjecture.     
 This conjecture can be justified by a thought experiment obtained by slight modification of  the folk no-global-charge  
  argument  presented above.  Indeed, instead of considering a global charge, consider a charge that is gauged 
  by a tiny  gauge coupling.  By making the gauge coupling sufficiently small, we can guarantee that the
  correction to the Schwarzschild metric due to the electric field is negligible.  Then the  effect  on the thermal radiation is
  negligible and non-emission of the weakly gauged charge continues essentially as in the global case.  This continues  until the point when it is too late to return the charge back.  This creates an inconsistency, which leads to a weak gravity conjecture.  
   
   Our  results show that there is a loophole in this reasoning, because it misses out the existence of a skyrmion type 
   hair created by the global baryonic charge. The strength of this hair is controlled by the global charge and persists 
   even in the limit of the zero gauge coupling.   It is this global hair, rather than the weak electric field sourced by it,  what   
   is responsible for a strong departure from Hawking's thermal regime.  Correspondingly, the conservation of 
   the charge is consistently accommodated, just as in the case of an ungauged global charge. 
       
   In conclusion, we discover that  irrespectively whether the baryon charge is a global or a very-weakly gauged symmetry, 
   once the black hole shrinks to a certain critical size it starts to reveal its baryonic content 
   in form of the skyrmion hair that grows inside out. After this point, irrespectively whether a black hole continues to shrink or simply stops, there is no inconsistency with the conservation of baryonic charge: 
   the baryonic content of a black hole is explicitly displayed by its skyrmion hair.  
   
   We must stress that there can certainly exist additional  sources of violation of  baryonic and other global  charges in 
   a microscopic theory.  The well-known example is violation of baryon and lepton numbers in grand unified theories. 
    Our arguments are in no contradiction with the existence of such sources. If such violation takes place, the skyrmion hair of a black hole can fade away over time.

\section{\label{sec:level1}Classical cross sections of a massless probe scalar field scattered by a skyrmion black hole}

We now choose two skyrmion black holes corresponding to two points of the parameter space discussed in section II: We consider the case $\alpha = 0.01$, $\beta = 0$, $x_h = x_h^{max, \alpha = 0.01, \beta = 0} = 0.1263$ (``CASE I") and the case $\alpha = 0.01$, $\beta = 0.5$ and $x_h = 0.116$ (``CASE II"). For both cases we study classical cross sections of a massless minimally-coupled probe scalar field $\Phi$ scattered by the skyrmion black hole. For simplicity we neglect possible non-gravitational interactions between the skyrmion and $\Phi$. It is however straightforward to take such interactions into account.

The motion of $\Phi$ in the background space-time of a skyrmion black hole can be described by the Klein-Gordon equation
\begin{equation}
\Box_g \Phi = 0 \, , 
\end{equation}
where $\Box_g$ is the d'Alambert operator in the space-time of a skyrmion black hole.

Using the expansion
\begin{equation}
\Phi = \frac{A(r)}{r} Y_{lm}(\theta, \phi) e^{-i W t} \,,
\end{equation}
where $Y_{lm}$ are the standard spherical harmonics, the radial part of the Klein-Gordon equation can be separated. After some algebra one can write this radial part as
\begin{equation}
\partial_{x*}^2A(x) + (w^2 - V_{eff}(x))A(x) = 0 \,,
\end{equation}
where
\begin{equation}
\partial_{x*} = N(x)h(x)\partial_x \,,
\end{equation}
\begin{equation}
V_{eff} =  N^2(x)h(x) \frac{l (l+1)}{x^2}+\frac{N(x)}{x}h(x)\partial_x\left(N(x)h(x)\right) \,,
\end{equation}
with
\begin{equation}
h(x) = \left(1-\frac{2m(x)}{x}\right)
\end{equation}
and a dimensionless frequency $w = W \left(e F_\pi\right)^{-1}$. In this form the radial part of the Klein-Gordon equation has the form of a Schr\"oedinger equation. Cross sections can therefore be studied by using methods of standard one dimensional quantum mechanical scattering theory. In the following subsections we study these cross sections both in CASE I and in CASE II for the two frequencies $w = 8$ and $w = 25$. In subsection A, we use a simple approximation to determine the scattering cross sections of high frequency waves at scattering angles $\theta \approx \pi$. In the subsections B and C we provide exact scattering and absorption cross sections which we obtained by performing a partial wave analysis. We use the same methods which have been extensively used in studies of classical scattering cross sections in the case of black holes from the Kerr Newman family \cite{futter, glam, *doran, *dolan, *crispino1, *crispino2, *crispino3, *crispino4, *crispino5, *crispino6, *crispino7}.

\subsection{\label{sec:level2}Geodesic motion and scattering cross sections using the glory approximation}

In the case of high frequencies ($w \gg 1$) the motion of a massless plane (scalar) wave can be approximated by the propagation of a ray moving on null geodesics \cite{futter, isaacson}. Geodesics in black hole backgrounds exhibit  interesting phenomena. In particular there exist impact parameters for which a particle propagating on geodesics moves once (or several times) around the black hole such that it is scattered back into the incoming direction \cite{mtw, darwin}. An asymptotic observer may therefore observe halos around the black hole. This phenomenon became well known as the ``glory phenomenon" \cite{matzner, *dewitt2}.

These two facts - the possibility to describe high frequency plane waves in terms of rays moving on geodesics and the existence of geodesic orbits going exactly once around the black hole - allow to calculate the scattering cross section for a scalar wave with frequency $w \gg 1$ in the glory regime (indeed for scattering angles $\theta \approx \pi$) by using the simple formula \cite{dewitt2}
\begin{equation}
\left(\frac{d\sigma}{d \Omega}\right)_{\theta \approx \pi} \approx 2 \pi w b_g^2 \left\lvert\frac{db}{d\theta}\right\rvert_{\theta = \pi} J_0^2\left(w b_g sin \theta\right)\,.
\end{equation}
Here $b$ is the dimensionless impact parameter (defined as $b = e F_\pi B$ with $B$ the impact parameter of the ray) and $b_g$ is the value of the dimensionless impact parameter for a geodesic ray moving exactly once around the black hole in such a way that it returns back to the same direction it came from. $J_0$ is a Bessel function of first kind.

\begin{figure}
\includegraphics[scale=0.3]{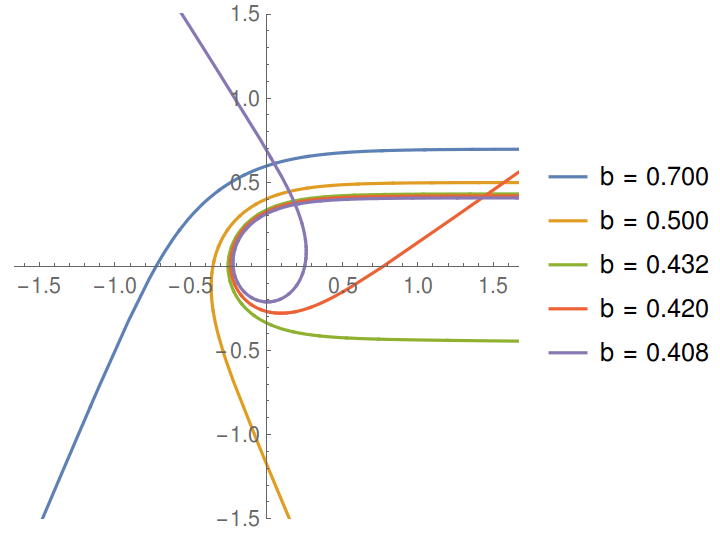}
\caption{Orbits for massless particles moving on null geodesics in the background of a skyrmion black hole of CASE I ($\alpha = 0.01$, $\beta = 0$ and $x_h = x_h^{max, \alpha = 0.01, \beta = 0} = 0.1263$) for different impact parameters with value $b$}
\label{fig:7}
\end{figure}

\begin{figure}
\includegraphics[scale=0.3]{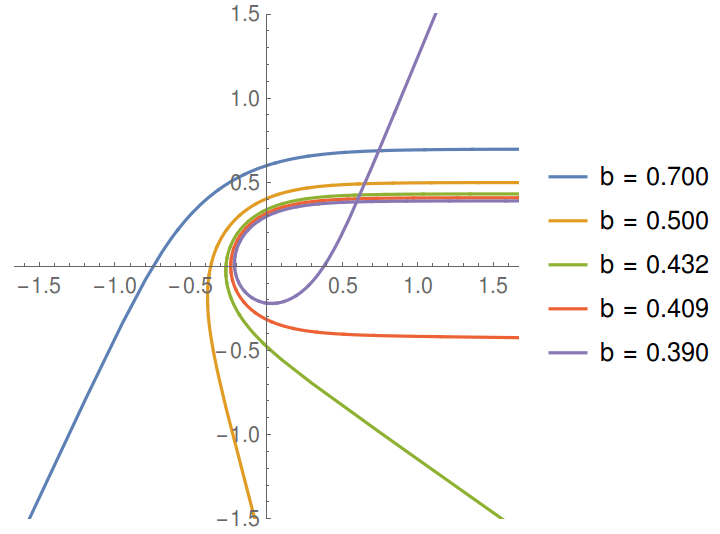}
\caption{Orbits for massless particles moving on null geodesics in the background of a skyrmion black hole of CASE II ($\alpha = 0.01$, $\beta = 0.5$ and $x_h = 0.116$) for different impact parameters with value $b$}
\label{fig:8}
\end{figure}

In Figures \ref{fig:7}
and \ref{fig:8}
we plot null geodesics for the two chosen cases of skyrmion black holes for different values of the impact parameter $b$. From these plots one can read off $b_g$: For CASE I we have $b_g = 0.432$ and for CASE II we have $b_g = 0.409$. In Figures 8, 9, 10 and 11 we plot the resulting scattering cross sections in the glory regime for the two cases both for $w = 8$ and for $w = 25$.

\subsection{\label{sec:level2}Scattering cross sections using a partial wave analysis}

We did a detailed partial wave analysis in order to determine the scattering cross sections for all scattering angles. Since $V_{eff}$ goes to $0$ for $x^* \rightarrow \infty$ we can write $A(x^*)$ for $x^* \rightarrow \infty$ as
\begin{equation}
A(x^*) = A^{(1)}_{wl} e^{-iwx^*} + A^{(2)}_{wl} e^{i w x^*}\, ,
\end{equation}
with two complex coefficients $A^{(1)}$ and $A^{(2)}$. At $x^* \rightarrow -\infty$ (indeed at $x \rightarrow x_h$) we choose the boundary condition
\begin{equation}
A(x^*) = A^{(3)}_{wl} e^{-iwx^*}\, ,
\end{equation}
with $A^{(3)}_{wl}$ another complex coefficient which satisfies $|{A^{(1)}_{wl}}|^2 = |{A^{(2)}_{wl}}|^2 + |{A^{(3)}_{wl}}|^2$. This boundary condition corresponds to a plane wave which comes from infinity and is purely ingoing. 

The differential scattering cross section for a scalar wave with frequency $w$ can be obtained as \cite{newton}
\begin{equation}
\frac{d\sigma}{d\Omega} = |h(\theta)|^2 \, ,
\end{equation}
where
\begin{equation}
h(\theta) = \frac{1}{2iw} \sum_{l=0}^{\infty} (2l +1)\left(e^{2i\delta_l(w)}-1\right)P_l(cos \theta)\, .
\end{equation}
Here $\delta_l$ are phase shifts which are defined as
\begin{equation}
e^{2 i \delta_l(w)} = (-1)^{l+1} \frac{A^{(2)}_{wl}}{A^{(1)}_{wl}} \, .
\end{equation}

We determined the phase shifts $\delta_l$ from $l=0$ up to a maximal value $l_{max}$. For calculating the sum we used the same ``method of reduced series" which has been invented in \cite{yennie} where it was successfully applied to the case of Coulomb scattering \footnote{This method has also been used in the scattering analysis of external fields on black holes within the Kerr Newman family \cite{futter, glam, *doran, *dolan, *crispino1, *crispino2, *crispino3, *crispino4, *crispino5, *crispino6, *crispino7}.}. This method ensures that we get reasonable results although we ignore in the sum all terms with $l > l_{max}$.

\begin{figure}
\includegraphics[scale=0.3]{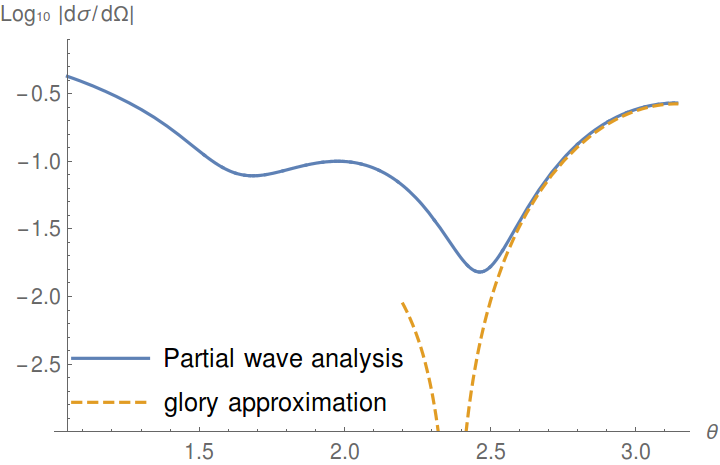}
\caption{Differential scattering cross section of a massless scalar wave with frequency $w = 8$ scattered by a skyrmion black hole of CASE I ($\alpha = 0.01$, $\beta = 0$ and $x_h = x_h^{max, \alpha = 0.01, \beta = 0} = 0.1263$)}
\label{fig:9}
\end{figure}

\begin{figure}
\includegraphics[scale=0.3]{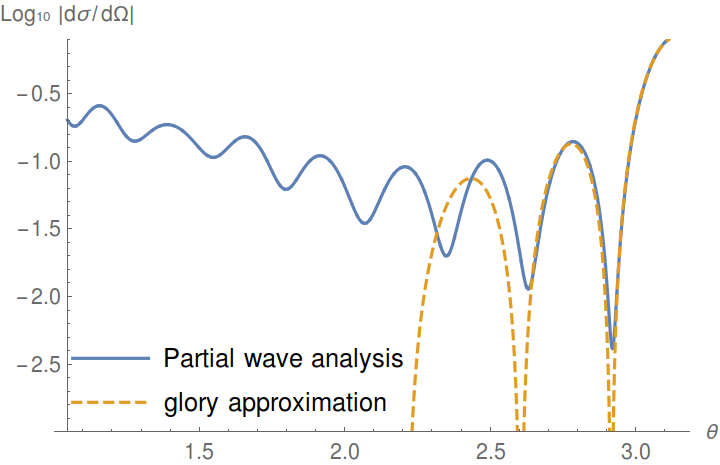}
\caption{Differential scattering cross section of a massless scalar wave with frequency $w = 25$ scattered by a skyrmion black hole of CASE I ($\alpha = 0.01$, $\beta = 0$ and $x_h = x_h^{max, \alpha = 0.01, \beta = 0} = 0.1263$)}
\label{fig:10}
\end{figure}

\begin{figure}
\includegraphics[scale=0.3]{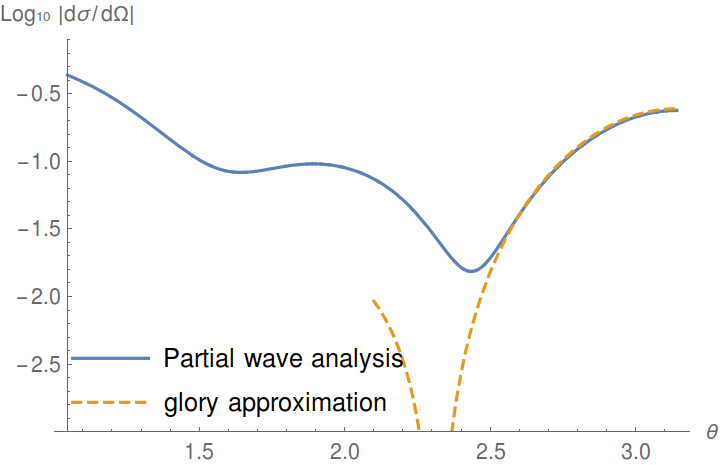}
\caption{Differential scattering cross section of a massless scalar wave with frequency $w = 8$ scattered by a skyrmion black hole of CASE II ($\alpha = 0.01$, $\beta = 0.5$ and $x_h = 0.116$)}
\label{fig:11}
\end{figure}

\begin{figure}
\includegraphics[scale=0.3]{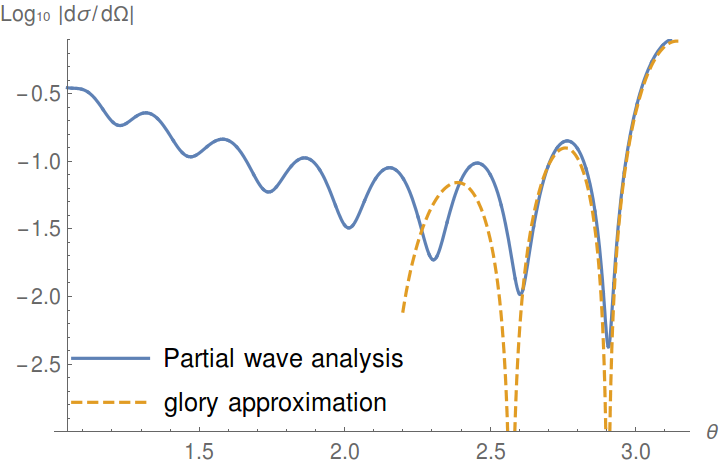}
\caption{Differential scattering cross section of a massless scalar wave with frequency $w = 25$ scattered by a skyrmion black hole of CASE II ($\alpha = 0.01$, $\beta = 0.5$ and $x_h =0.116$)}
\label{fig:12}
\end{figure}

We plot the results in Figure \ref{fig:9}
, Figure \ref{fig:10}
, Figure \ref{fig:11}
and Figure \ref{fig:12}
. From the plots we see that the cross sections for CASE I and CASE II only very slightly differ for both frequencies.\footnote{If one looks carefully at the plots one realizes that the characteristic peaks of the cross section in CASE I are positioned at slightly bigger angles when compared to the peaks of the cross section in CASE II.} This was already expected from the analysis in subsection A where it was shown that $b_g$ is only slightly different in CASE I and CASE II.

\subsection{\label{sec:level2}Absorption cross sections using a partial wave analysis}

From the same partial wave analysis the partial absorption cross sections $\sigma_l$ for a scalar wave with frequency $w$ can be obtained as \cite{newton}
\begin{equation}
\sigma_l = \frac{\pi}{w^2}(2l+1)\left(1-\left\lvert\frac{A^{(2)}_{wl}}{A^{(1)}_{wl}}\right\rvert^2\right) \, .
\end{equation}
The total absorption cross section $\sigma_{abs}$ for a given frequency $w$ can be obtained as the sum over all $\sigma_l$. Since from doing the partial wave analysis we found that there exists - for the accuracy we achieved in our numerical calculations - a $\tilde{l}$ such that $\sigma_l = 0$ for all $l > \tilde{l}$, this sum is finite. For CASE I and $w = 8$ we found $\sigma_{abs} = 0.48$. For CASE I and $w = 25$ we found $\sigma_{abs} = 0.51$. For CASE II and $w = 8$ we found $\sigma_{abs} = 0.45$. For CASE II and $w = 25$ we found $\sigma_{abs} = 0.46$.

\section{\label{sec:level1}Comparison of the scattering cross sections
for black holes with and without skyrmion hair.}

One way of investigating how the presence of the skyrmion black hole hair manifests itself in the scattering cross sections is to compare the scattering cross section of the scalar field scattered by the skyrmion black hole with the scattering cross section of the same scalar field scattered by a Schwarzschild black hole. We do this for CASE I and CASE II.

\begin{figure}
\includegraphics[scale=0.3]{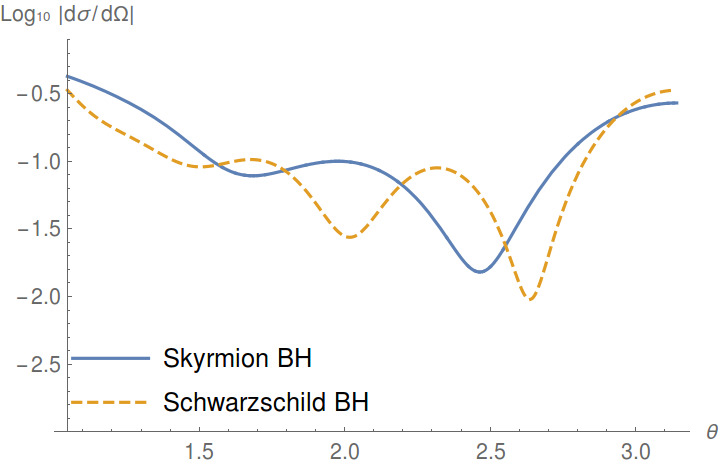}
\caption{Differential scattering cross section of a massless scalar wave with frequency $w = 8$ scattered by a skyrmion black hole of CASE I ($\alpha = 0.01$, $\beta = 0$, $x_h = x_h^{max, \alpha = 0.01, \beta = 0} = 0.1263$ and $m_{ADM} = 0.111$) and scattering cross section of the same scalar wave scattered by a Schwarzschild black hole with mass $M = 0.111$}
\label{fig:13}
\end{figure}

\begin{figure}
\includegraphics[scale=0.3]{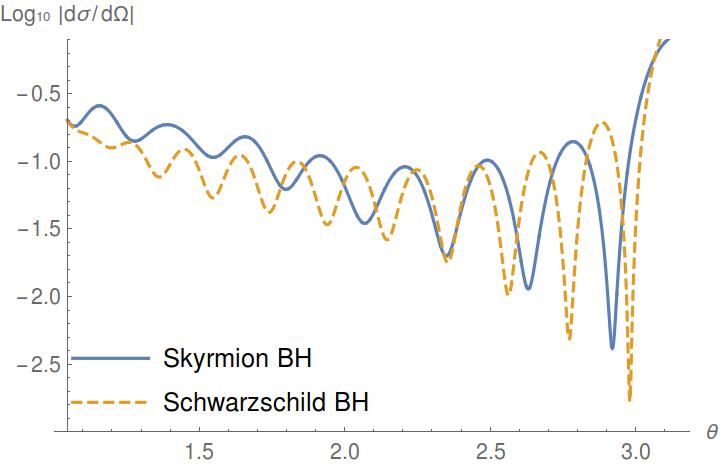}
\caption{Differential scattering cross section of a massless scalar wave with frequency $w = 25$ scattered by a skyrmion black hole of CASE I ($\alpha = 0.01$, $\beta = 0$, $x_h = x_h^{max, \alpha = 0.01, \beta = 0} = 0.1263$ and $m_{ADM} = 0.111$) and scattering cross section of the same scalar wave scattered by a Schwarzschild black hole with mass $M=0.111$}
\label{fig:14}
\end{figure}

In the first case we compare the differential scattering cross section of the scalar field scattered by a skyrmion black hole of CASE I with the differential scattering cross section of the same scalar field scattered by a Schwarzschild black hole which has the same ADM mass as the skyrmion black hole of CASE I. We plot the results in Figure \ref{fig:13}
and Figure \ref{fig:14}.
From the plots we can see that for both frequencies the presence of the hair ``shifts" the characteristic peaks in the differential scattering cross section to smaller scattering angles. This comes from the fact that the hair ``changes" the near horizon geometry of the black hole and the scalar field therefore ``sees" a ``mass" which is smaller than the ADM mass when moving in the near horizon zone. (It is known that in the case of a Schwarzschild black hole with mass $M_1$ the characteristic peaks appear at smaller scattering angles when compared to a Schwarzschild black hole with mass $M_2 > M_1$ \cite{futter}. Therefore it is plausible that for a skyrmion black hole the peaks are ``shifted" to smaller scattering angles and not, for example, to bigger scattering angles when compared to a Schwarzschild black hole with the same ADM mass.)

\begin{figure}
\includegraphics[scale=0.3]{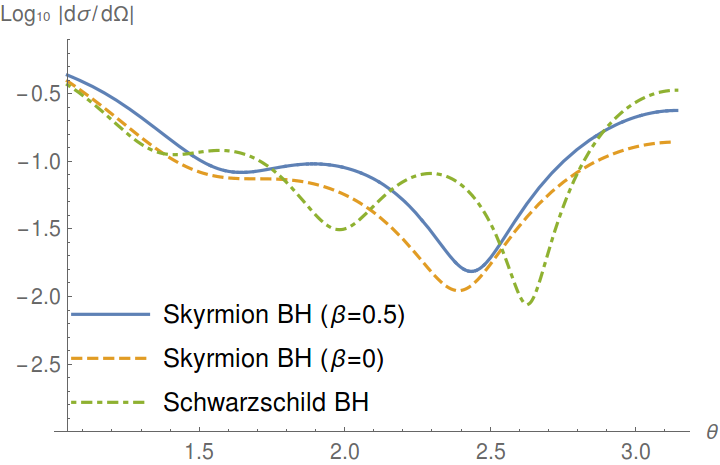}
\caption{Differential scattering cross section of a massless scalar wave with frequency $w = 8$ scattered by a skyrmion black hole of CASE II ($\alpha = 0.01$, $\beta = 0.5$, $x_h = 0.116$ and $m_{ADM} = 0.108$), scattering cross section of the same scalar wave scattered by a skyrmion black hole with $\alpha = 0.01$ and the same ADM mass $m_{ADM} = 0.108$ but with $\beta = 0$ and scattering cross section of the same scalar wave scattered by a Schwarzschild black hole with mass $M=0.108$}
\label{fig:15}
\end{figure}

\begin{figure}
\includegraphics[scale=0.3]{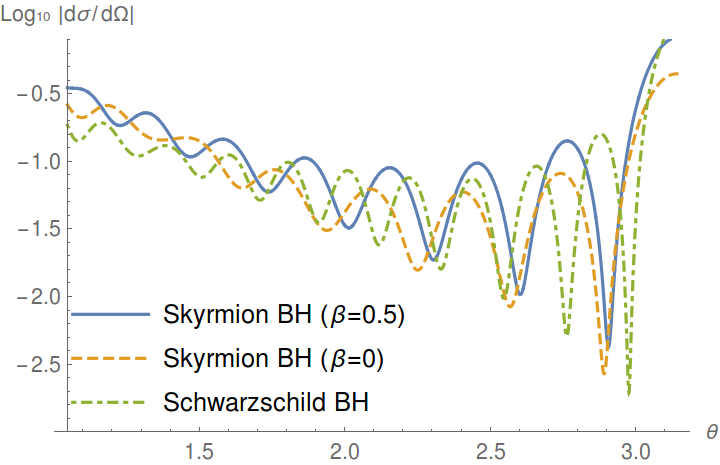}
\caption{Differential scattering cross section of a massless scalar wave with frequency $w = 25$ scattered by a skyrmion black hole of CASE II ($\alpha = 0.01$, $\beta = 0.5$, $x_h = 0.116$ and $m_{ADM} = 0.108$), scattering cross section of the same scalar wave scattered by a skyrmion black hole with $\alpha = 0.01$ and the same ADM mass $m_{ADM} = 0.108$ but with $\beta = 0$ and scattering cross section of the same scalar wave scattered by a Schwarzschild black hole with mass $M=0.108$}
\label{fig:16}
\end{figure}

In the second case we compare the differential scattering cross section of the scalar field scattered by a skyrmion black hole of CASE II both with the differential scattering cross section of the same scalar field scattered by a Schwarzschild black hole with same ADM mass and with the differential scattering cross section of the same scalar field scattered by a skyrmion black hole with the same ADM mass, with $\alpha = 0.01$ but (in contrast to CASE II) with $\beta = 0$. We plot the results in Figure \ref{fig:15}
 and Figure \ref{fig:16}
. As in the first case we see that the presence of the "hair" ``shifts" the peaks to smaller scattering angles. From comparing CASE II to the skyrmion black hole with same $\alpha$, same ADM mass but with pions which are massless we see that the pion masses "shift" the peaks to bigger scattering angles when compared to the massless pion case. \\

\section{\label{sec:level1}Summary and outlook\protect\\}

In this work we have studied some important implications of classical skyrmion black hole hair.
\newline

First, we showed that the  existence of skyrmion hair  leads to a loophole in the folk theorems which are claiming an inconsistency of global symmetries 
with semi-classical black hole physics. 
In order to study the situation we have carefully analyzed the Gedankenexperiment that  one uses in standard 
``proofs" of folk theorems.  In this  Gedankenexperiment  a black hole, which has swallowed a baryon,
evaporates via Hawking radiation.  The conclusion reached in the standard treatment is that the baryon number cannot be conserved in this process.   However, this reasoning assumes that  no classical hair - that could carry information about the baryonic charge - can exist at any stage of the black hole evaporation.  
  We have argued that in the light of the two facts, the existence of skyrmion black holes and the skyrmion/baryon correspondence, the above assumption is unjustified. 
  We have performed a refined analysis in which the new information is taken into the account and reached a conclusion 
which seriously challenges the folk theorem.  
In our version of the thought experiment the originally-swallowed baryonic charge re-surfaces in form 
of the skyrmion hair, once the black hole shrinks down to a scale $L$ determined by the baryonic charge.  

We argued that the skyrmion black holes are natural candidates to describe black holes with baryon number, 
at stages when the size of the event horizon of the black hole, $r_h$, becomes of order or smaller than the typical length of the skyrmion/baryon, $L$.  Thus, in this  scenario the  skyrmion/baryon  simply reappears once the black hole shrinks down to the  size $L$ and the black hole shrinks further down to Planck size inside the skyrmion/baryon. In this sense skyrmions/baryons get recycled and baryon number is conserved in systems with black holes.

 For our Gedankenexperiment we have chosen a setup of large-$N_C$ QCD, with massless quark flavors.   
 By varying $N_C$ and the QCD-length, we could on the one hand make the size of the baryon/skyrmion ($L$)
 arbitrarily large, and on the other hand, keep the mapping between the skyrmion topological current and 
 the baryon number current arbitrarily accurate.  Hence, we can create a situation in which the critical 
 size of the black hole event horizon at which the skyrmion hair emerges (around  $r_h \sim L$)  can be made arbitrarily larger than the Planck length.  
 
  Our analysis also  exposes a loophole in the reasoning leading to a weak gravity conjecture  \cite{weak1, *weak2}, according to which 
  the gauge coupling cannot be arbitrarily weak (when measured in units of gravitational coupling) in order to be compatible with  semi-classical black hole physics.  
     This conclusion is also based on the assumption of the absence 
  of any classical hair in the limit of zero gauge coupling.  However, as we argued, it is not valid in the light of the 
  existence of skyrmion black hole hair.   Performing the Gedankenexperiment in the presence of a 
  gauge field coupled to quark baryonic current infinitesimally-weakly, we show that there is no evidence of any inconsistency  
 with semi-classical black hole physics.  
 
  Of course, in this work we are not proving that there cannot exist an independent fundamental principle that  excludes  
 global and/or very weakly-gauged symmetries.  This has to be investigated additionally on case-by-case basis.  For example, in high-dimensional theories in which the four-dimensional gauge symmetries originate from the isometries of compact manifolds, the weak-gravity conjecture is built-in.  Similarly, in weakly-coupled string theory, where both the gauge bosons and a graviton originate as string modes, their relative coupling strengths are related and cannot be arbitrarily small.    
  However, what we are pointing out in the present paper is a serious loophole in the  standard arguments claiming that such symmetries are necessarily  inconsistent with the semi-classical black hole physics.   
  In other words, evidently, whether a global symmetry, such as, e.g.,  baryon number, is a good symmetry or not, is decided by a fundamental theory and not by the semi-classical black hole physics.

   As a byproduct we have shown that in order for QCD to remain a well-defined theory in the presence of gravity, the number of colors $N_C$ and the QCD length $L$ must obey the relation (\ref{species}). 
  This relation represents a particular case of the black hole bound on the number of species \cite{giabound, giabound2, giabound3} applied 
  to QCD with $N_C$ colors.  A violation of the bound (\ref{species}) implies that the length-scale of strong quantum gravity exceeds the QCD-length and thus it stops to be a valid theory of either quarks and gluons or mesons, baryons and glueballs. 
 Remarkably, 
 this phenomenon manifests itself already within the semi-classical treatment.  We have observed that whenever the bound 
 (\ref{species}) is violated, the states with quark $N_C$-iality become 
 more compact than their gravitational radii, i.e., they 
 become black holes and theory no longer contains any baryons. \\

Second, we studied cross sections of a massless minimally-coupled probe scalar field scattered by a black hole with classical ``hair". As a working example we used the skyrmion black holes. We considered two explicit examples of skyrmion black holes and studied in these two cases classical absorption and scattering cross sections both for a scalar wave with frequency $w = 8$ and a scalar wave with frequency $w = 25$. We compared the differential scattering cross sections with the differential scattering cross sections of the same scalar waves scattered by a Schwarzschild black hole which has the same ADM mass as the skyrmion black hole. We found an order-one difference both in the case of massless and in the case of massive pions, which manifests itself in a shift of the characteristic peaks in the differential scattering cross sections to different scattering angles. This effect  comes from the fact that the presence of the ``hair" changes the near-horizon geometry of the black hole when compared to a Schwarzschild black hole with same ADM mass. (We did not take into account possible non-gravitational interactions between the scalar field and the skyrmion).
\newline

Finally some comments are in order. 
First, despite the fact that  we considered a simplified model, in which the low-energy effective theory of QCD is approximated by a chiral theory of pions,  we believe that our main qualitative result - that the peaks in the differential scattering cross sections are shifted to smaller scattering angles when compared to a Schwarzschild black hole with the same ADM mass - will also show up in more realistic models because this effect comes only from the fact that the near-horizon geometry is changed by the presence of the ``hair" - a fact which we expect to be there for more realistic models of baryons. The same is expected to hold for similar configurations as the skyrmion black holes, which can also be interesting for astrophysical situations, such as the recently discovered ``hairy" Kerr black holes \cite{herd, *herd2, *herd3, *herd4, *herd5}\footnote{Whether these hairy Kerr black holes are truly relevant for astrophysical situations, is an open question. From the theoretical point of view the stability (against linear perturbations) seems to be a necessary condition for these black holes to be relevant for astrophysics. To our knowledge the fact of stability has not been established so far. }. Second, the hair that we have focused on in our analysis was purely classical. We considered a classical model for the skyrmion and we used classical theory of gravity to describe the black hole.  As we have seen, the skyrmion black holes only exist as classical solutions of the Einstein equations as long as the sum of the black hole event horizon $r_h$  and the gravitational radius of the skyrmion $L_g$ is not  bigger than the typical length of the skyrmion $L$ (formally this fact manifests itself in the presence of a maximal possible horizon size $r_h^{max, \alpha, \beta}$). As soon as the sum of  $r_{h}$ and $L_g$ becomes bigger than $L$,  from the point of view of an external classical observer the system is pure Schwarzschild. Capturing the effects of the baryonic/skyrmion  charge 
of a black hole in this case probably requires a quantum resolution of the black hole state. 
 In this respect,  it is interesting to note that the re-appearance of the classical skyrmion hair  below the critical scale, set  by the condition 
$r_h + L_g \sim L$, can be taken as supporting evidence for the existence of the quantum baryonic hair of a black hole \cite{dvaligomez}. 
Indeed, the two facts, that for $r_h + L_g \gg L$ there is no classical skyrmion hair and that it resurfaces in the regime 
$r_h + L_g  \ll L$, probably indicates that the information about the baryon/skyrmion charge must be present also in the domain $r_h + L_g  \gg L$ in
some form and the quantum hair of the type \cite{dvaligomez} is a viable candidate for carrying such information.   This is fully compatible with the fact that 
 at all times the baryon/skyrmion charge of a black hole is encoded  
 in the boundary integral (\ref{Intboundary}), which is topological in nature.

\begin{acknowledgments}

 It is pleasure to thank Cesar Gomez  for valuable discussion on the physical meaning of baryonic hair 
 within large-$N_C$ QCD coupled to gravity, as well as on other aspects of this work. We thank Silke Britzen for general discussions on possible astrophysical implications of black hole baryonic hair.
We are grateful to  A. Averin, D. Flassig, A. Helou, S. Hofmann, O. Iarygina, F. Niedermann, T. Rug, D. Sarkar, M. Shifman and N. Wintergerst for discussions and/or useful comments. 
 
The work of G. D. was supported by Humboldt Foundation under Alexander von Humboldt Professorship, by European Commission under ERC Advanced Grant 339169 ``Selfcompletion", by DFG SFB/TRR 33 ``The Dark Universe" and by the DFG cluster of excellence EXC 153 ``Origin and Structure of the Universe". The work of A. G. was supported by the DFG cluster of excellence EXC 153 ``Origin and Structure of the Universe".

\end{acknowledgments}

\appendix

\section{Field equations in the Einstein skyrme system}

For completeness we provide the field equations we used in order to solve the Einstein skyrme system. For the case of $\beta = 0$ these equations are also given for example in \cite{bizon}. When comparing the field equations to the field equations in other works like \cite{shiiki} one has to take into account that different conventions for the signature of the metric, 
different conventions for the pion decay constant $F_\pi$ and/or different conventions for the skyrme Lagrangian $\mathcal{L}_S$ are often used.

One starting point to obtain the field equations is the action
\begin{equation}
S = \int \sqrt{-g} \left(-\frac{1}{16\pi G_N} R + \mathcal{L}_S\right) d^4x \, ,
\end{equation}

where $R$ is the Ricci scalar. For the metric we use the ansatz (\ref{metricansatz}). For the pion field we use the hedgehog ansatz (\ref{hedgehog}). Varying $S$ with respect to the metric gives the Einstein equations (\ref{einequ})
\begin{equation}
G_{\mu \nu} = 8 \pi G_N T^S_{\mu \nu} \, .
\end{equation}
Varying $S$ with respect to $F$ gives the equation
\begin{equation}
\begin{split}
& \partial_x\left(\left(x^2+2\mathrm{sin}^2F\right)N(x)h(x)\partial_xF\right)  \\
& = N(x)\left\{\mathrm{sin}2F\left(1+h(x)\left(\partial_xF\right)^2+\frac{\mathrm{sin}^2F}{x^2}\right)+\beta^2x^2\mathrm{sin}F\right\} \, .
\end{split}
\end{equation}

Here we introduced dimensionless quantities $x = e F_\pi r$ and $m(x) = e F_\pi G_N M(r)$ and the function $h(x)$ which is defined as
\begin{equation}
h(x) = 1-\frac{2m(x)}{x} \, .
\end{equation}

In order to solve the whole system two non-vanishing independent components of the Einstein equations are sufficient. The temporal and radial components can be written as
\begin{equation}
\begin{split}
\partial_xm & = \alpha \Big\{\frac{x^2}{2}h(x)\left(\partial_xF\right)^2 \\
&+ \mathrm{sin}^2F +\mathrm{sin}^2F\left(h(x)\left(\partial_xF\right)^2 + \frac{\mathrm{sin}^2F}{2x^2}\right) \\
&- \frac{1}{2}\beta^2x^2\left(2\mathrm{cos}F - 2\right)\Big\}
\end{split}
\end{equation}
\begin{equation}
\partial_xN = \alpha \left(x+\frac{2}{x}\mathrm{sin}^2F\right)N(x)\left(\partial_xF\right)^2 \, .
\end{equation}
Note that the right hand side of the Einstein equations scales linear in $\alpha$.

\section{A gas of skyrmion black holes}

The skyrmion black hole configurations considered in the main part of this work are configurations which arise as solutions of the Einstein equations when the energy momentum tensor of one skyrmion (corresponding to a skyrmion in flat spacetime with $B=1$) is taken as a source in the equations. 

 The solutions can be generalized to a multi-skyrmion situation.  In the case when the skyrmions are well-separated (by distances $\gg L$) the solution is particularly simple and represents a gas of skyrmion black holes. 
   
 For $N_B$ skyrmions (each with $B=1$) the Lagrangian can be taken as
\begin{equation}
\mathcal{L}_S^{(gas)}\left(U^{(1)}, U^{(2)}, ..., U^{(N_B)}\right) = \sum_{n=1}^{N_B} \mathcal{L}_S(U^{(n)}) \, .
\end{equation}
Here $\mathcal{L}_S(U^{(n)})$ is the skyrme Lagrangian introduced in (\ref{lagrangian}) for the matrix field $U^{(n)}$ of a single skyrmion.  Solitonic configurations can be obtained in complete analogy to the case with single  skyrmion as discussed in section IIB if we make for each $U^{(n)}$ the hedgehog ansatz
\begin{equation}
\frac{\pi_a^{(n)}}{F_\pi} = F^{(n)}(r-r_n)n_a
\end{equation}
and minimize the energy functional of the gas using the boundary conditions
\begin{equation}
F^{(n)}(r_n) = \pi, F^{(n)}(\infty) = 0 \, .
\end{equation}
The solution profile functions $F^{(n)}$ are analogous to the profile function in the case of only one skyrmion as discussed in section IIB. The solution describes a gas of $N_B$ non-interacting skyrmions located at the space points $r_1, ..., r_{N_B}$.

Note that the same solution appears if - at the very beginning - one starts with a product ansatz
\begin{equation}
\mathcal{L}_S^{(gas)} = \mathcal{L}_S\left(U^{(1)}(r-r_1)U^{(2)}(r-r_2)...U^{(N_B)}(r-r_{N_B})\right)
\end{equation}
and at the same time assumes that the distances between the individual skyrmions are so large that the overlaps of the solution profile functions of the individual skyrmions in the gas can be safely neglected. This approach was taken in \cite{skyrmgas, *skyrmgas2}.

One can study the skyrme gas coupled to gravity by using the energy momentum tensor corresponding to $\mathcal{L}_S^{(gas)}$ (defined as in (\ref{energymomentum}) with $\mathcal{L}_S$ replaced by $\mathcal{L}_S^{(gas)}$) as a source in Einstein's equations. This energy momentum tensor now takes the form
\begin{equation}
T_{\mu \nu}^S\left(U^{(1)}, U^{(2)}, ..., U^{(N_B)}\right) = \sum_{n=1}^{N_B}T_{\mu \nu}^S\left(U^{(n)}(r-r_n)\right)\, .
\end{equation}

In this case the Einstein equations can be solved for each summand separately and for each summand the solution is the same as the solution of the Einstein equations with the energy momentum tensor of only one skyrmion taken as a source (discussed in section IIC). Since, as discussed in section IIC, solutions for each summand exist only within a certain parameter space, in particular only allowing for solutions with event horizon located inside a skyrmion, we now correspondingly get a solution which describes a skyrmion gas with $N_B$ skyrmions (located at $r_1, ..., r_{N_B}$) all having a black hole event horizon located inside.

\section{Constraint on Semi-Classical Gravity  from the Number of Colors}  

 In this appendix we would like to show how the number of colors restricts the domain of validity of QCD via 
 the bound (\ref{species}).  Beyond this domain QCD simply stops existence as a well-defined theory 
 that can be treated without the knowledge of strong quantum gravity.  The chiral Lagrangian can formally be written  down at low energies, but  it is no longer a trustable theory of strong interactions. In particular, there is no possibility of  consistently including baryons in the theory. 

 Let us first discuss the physical meaning of the bound (\ref{species}).  This bound comes from the general black hole bound on the number of particle species that can be consistently treated within the semi-classical gravity at weak-coupling \cite{giabound, giabound2, giabound3}. 
 That is,  the number of particle species for which the strong-quantum gravity effects can be ignored. 
 Let the length scale at which quantum gravity becomes strong be $L_{QG}$.  This means  that for 
 processes with momentum-transfer above  the energy scale $M_{QG} \equiv {\hbar \over L_{QM}}$ quantum  gravitational coupling is strong.  The effective theory in which such effects can be ignored can only be defined at distances larger 
 than $L_{QG}$.  Equivalently, $L_{QG}$ sets the resolution scale of any measurement reliably performed in semi-classical gravity.   The scale  $L_{QG}$ is not unique and depends on the theory. For example, in pure Einstein gravity with no additional particles with masses below the Planck mass, the scale $L_{QG}$ coincides with the Planck length, $L_P$.   
 But, in general, this is not the case. 
 Let $N$ be the number of independent particle species that can be consistently treated in the low energy theory of semi-classical gravity.  Then, black hole bound on the species scale tells us that for $N \gg 1$, the scale $L_{QG}$ is related with the Planck length in the following way 
 \begin{equation}
    L_{QG} \,  =  \, \sqrt{N} L_P\, .
 \label{scale} 
 \end{equation}
 Thus, in theory with many particle species, the length-scale beyond which semi-classical gravity cannot be trusted becomes larger.  In \cite{giabound, giabound2, giabound3} several versions of the proofs of this bound were given. The essence of these proofs is to show - by designing thought 
 experiments - that a black hole semi-classicality cannot be maintained beyond the scale $L_{QG}$ given by (\ref{scale}). 
 
   Let us directly re-derive this bound for the current situation of large-$N_C$ QCD. We shall assume that the quark flavors are only two and are massless.  Let us consider a large neutral black hole with the initial Hawking temperature 
  much below the QCD scale $\hbar/L$.  Until the black hole temperature reaches the QCD scale, the black hole 
  can only evaporate into pions.  The glueballs and baryons are much heavier and their production rate is negligible due to Boltzmann suppression.   The way we can understand the evaporation process is to think that in each emission act the black hole produces either a quark or an anti-quark which then picks up a partner from the quark see and hadronizes. The effective number of emission channels scales as $N_C$.  (Therefore, in the situation at hand, the effective number of species $N$, relevant for black hole evaporation, is $N_C$.) Taking this factor into the account we can evaluate the time-change of the black hole temperature as 
  \begin{equation}
  {\dot{T} \over T^2} \, = \, N_C \left ({T \over M_P }\right )^2 \, . 
  \label{colorT}
  \end{equation} 
   The left hand side of this equation represents a semi-classicality parameter. In particular,  it measures the deviation from 
   the thermal emission.  Therefore,  once this parameter becomes order one, 
   semi-classical gravity can no longer be trusted. It is clear that this breakdown of semi-classicality happens around temperatures given by $T = {\hbar  \over L_{QG}}$, where the scale   $L_{QG}$ is given by (\ref{scale}). Obviously, in order for QCD to be a well-defined theory 
    the quantum gravity length $L_{QG}$ must be shorter than the QCD length $L$.  This requirement  gives the bound on the number of colors 
    given by (\ref{species}). If this bound is violated, QCD simply is no longer a theory that could even be defined without the knowledge of the strong quantum gravity domain.  \\
   
    It is very interesting that even if we  completely ignore the above knowledge and blindly continue to apply the semi-classical gravity 
    beyond the bound (\ref{species}),  theory will not allow us to go very far.
     Let us illustrate this point. Let us assume that we take $N_C \gg N_*$.  We immediately realize that 
     the gravitational radius of a baryon exceeds  the QCD length, and thus, the non-gravitational size of a baryon. 
    In other words, a baryon only exist in form of a black hole.  In this situation it is unclear how to even define a baryon. 
 Normally, for this definition we need to be able to prescribe to a given state a quark $N_C$-iality.  Which is clearly 
 impossible if any such state is a classical black hole!      
     
\begin{figure}
\includegraphics[scale=0.2]{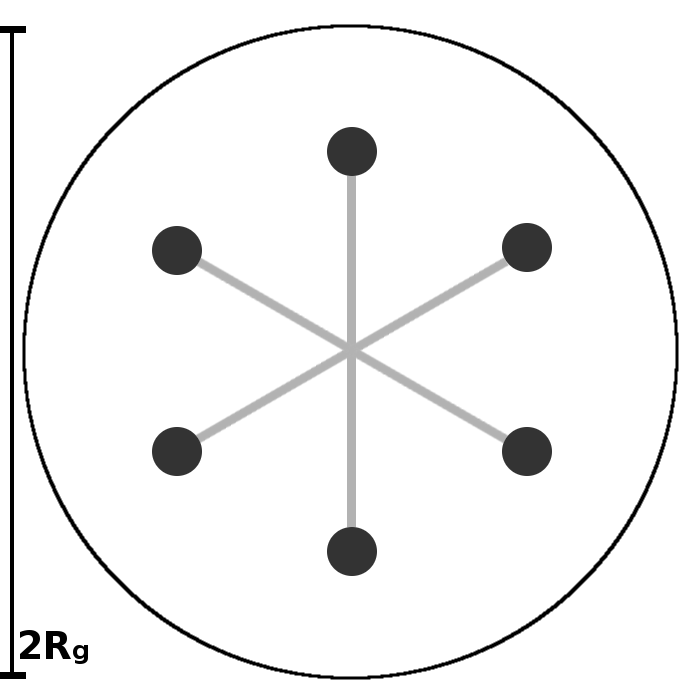}
\caption{Illustration of a state with $N_C = 6$ quarks attached to a single junction by QCD flux tubes in the case when the length of each string $R$ is smaller than the gravitational radius $R_g$ of the state}
\label{fig:17}
\end{figure}

        Furthermore,  it is easy to understand that  no 
     $N_C$-quark states can be defined that would be larger than its own  gravitational radius. Not even an excited state. 
     Indeed, consider a state  with $N_C$ quarks that by QCD flux tubes are attached to a single junction (see Fig. \ref{fig:17} for an illustration). 
     In the absence of gravity,  such a configuration in its lowest energy state describes a baryon.   
    In the present situation, as we saw above, a non-excited baryon is a black hole.     
      Let us ask whether by pulling the quarks apart we can make the size of this configuration bigger than its gravitational radius. 
      If we succeed, we will be able to identify at least some excited states with quark $N_C$-iality.  
      This however is not possible. 

    Let the length 
     of each string be $R \gg L$.  Since, the string tension is set by the QCD scale, the mass of each string is 
     $R {\hbar^2 \over L^2}$ and the total mass scales as $M \sim N_CR {\hbar^2 \over L^2}$.  Evaluating the gravitational radius of this configuration we  get, 
     \begin{equation}
     R_g \, \sim \, R{N_C \over N_*} \, , 
     \label{RgN}
     \end{equation}
    which exceeds $R$  whenever the bound (\ref{species}) is violated.  Not surprisingly,  for $R=L$ and $R_g=L_g$,  the equation
    (\ref{RgN})  reproduces (\ref{ratioL}).

         Thus, the theory which is much smarter than us, is telling us that 
   even if we ignore the quantum gravity bound, the resulting theory will not  allow existence of  any asymptotic (or even excited) state that can be treated as baryon.  The notion of baryons simply stops to exist in this theory.  This is the way the semi-classical theory is telling us that by violating the bound (\ref{species}) we entered the domain in which neither semi-classical gravity nor QCD are any longer well-defined theories.

\bibliography{apssamp}

\end{document}